\documentclass[journal]{IEEEtran}
\usepackage{lipsum}
\usepackage[utf8]{inputenc}
\usepackage{amsmath,amsfonts,amssymb,bm,balance}
\usepackage{graphicx}
\usepackage{subcaption}
\usepackage{caption}
\usepackage{xcolor}
\usepackage{mathtools}
\usepackage{comment}
\usepackage{textcomp}

\usepackage{multirow}
\usepackage{epstopdf}
\usepackage{cite}
\usepackage{arydshln}
\usepackage{blkarray, multirow}
\usepackage{arydshln}
\usepackage{tikz,pgfplots}
\usetikzlibrary{arrows, arrows.meta, backgrounds, calc, intersections, decorations.markings, decorations.pathreplacing, positioning, shapes}
\pgfplotsset{compat=newest}

\newcommand{\bi}{\begin{itemize}}
\newcommand{\ei}{\end{itemize}}

\newcommand{\be}{\begin{enumerate}}
\newcommand{\ee}{\end{enumerate}}
\newcommand{\bd}{\begin{description}}
\newcommand{\ed}{\end{description}}
\newcommand{\bc}{\begin{center}}
\newcommand{\ec}{\end{center}}
\newcommand{\bt}{\begin{tabbing}}
\newcommand{\et}{\end{tabbing}}
\newcommand{\bfig}{\begin{figure}}
\newcommand{\efig}{\end{figure}}
\newcommand{\beq}{\begin{equation}}
\newcommand{\beqarr}{\begin{eqnarray}}
\newcommand{\beqarrn}{\begin{eqnarray*}}
\newcommand{\eeq}{\end{equation}}
\newcommand{\eeqarr}{\end{eqnarray}}
\newcommand{\eeqarrn}{\end{eqnarray*}}
\newcommand{\bflr}{\begin{flushright}\vspace{-0.2in}}
\newcommand{\eflr}{\end{flushright}}
\newcommand{\bsub}{\begin{subequations}}
\newcommand{\esub}{\end{subequations}}
\newcommand{\barr}{\begin{array}}
\newcommand{\earr}{\end{array}}
\newcommand{\nn}{\nonumber}
\def\binom#1#2{\left( \!  \barr{c} #1 \\ #2 \earr \!  \right)}

\def\undb#1{\mbox{\bf{#1}}}

\usepackage{titlesec}
\let\subparagraph\relax
\titlespacing{\section}{0pt}{6pt plus 2pt minus 1pt}{4pt plus 1pt minus 1pt} 
\titlespacing{\subsection}{0pt}{5pt plus 2pt minus 1pt}{3pt plus 1pt minus 1pt} 
\begin{document}

\title{Outage Probability Analysis of MRC-Based Fluid Antenna Systems under Rician Fading}
\author{Soumya~P.~Dash,~\IEEEmembership{Senior Member,~IEEE}, Tummi Ganesh, and Italo~Atzeni,~\IEEEmembership{Senior Member,~IEEE}
\thanks{S. P. Dash is with the School of Electrical and Computer Sciences, Indian Institute of Technology Bhubaneswar, Argul, Khordha, Odisha, 752051 India, and with the Centre for Wireless Communications, University of Oulu, 90570 Oulu, Finland (e-mail: soumyapdashiitbbs@gmail.com).}
\thanks{T.~Ganesh is with the School of Electrical and Computer Sciences, Indian Institute of Technology Bhubaneswar, Argul, Khordha, 752050 India (e-mail: 24sp06008@iitbbs.ac.in).}
\thanks{I. Atzeni is with the Centre for Wireless Communications, University of Oulu, 90570 Oulu, Finland (e-mail: italo.atzeni@oulu.fi).}
\thanks{The work of I. Atzeni was supported by the Research Council of Finland (336449 Profi6, 348396 HIGH-6G, and 369116 6G~Flagship).}
}
\maketitle
\begin{abstract}
This paper investigates a fluid antenna system (FAS) where a single-antenna transmitter communicates with a receiver equipped with a fluid antenna (FA) over Rician fading channels. Two channel models are considered to incorporate the correlation among the ports into the fading gains, namely: i) the widely used {\em physical reference port model}, implemented by considering the first physical port as the reference port; and ii) the {\em virtual reference port model}, considering a common port correlation coefficient and implemented by assuming the presence of a virtual reference port. Assuming that multiple ports among the $M$ available FA ports can be activated, the receiver selects the best $K$ with the highest instantaneous signal-to-noise ratio (SNR) and combines the received signals at the selected ports using maximum ratio combining. For both port models, the statistics of the post-combining SNR are derived using a characteristic function approach, which allows to analyze the outage probabilities (OPs) of the FAS. Additional closed-form lower bounds on the OPs and the asymptotic OPs at high SNR are derived, revealing the diversity order of the FAS to be $M$. Numerical results validate the analytical framework and demonstrate the interplay of key system parameters on the performance of the considered MRC-based FAS. Specifically, the performance difference between the two reference port models becomes increasingly evident as $M$, $K$, and the average SNR of the FAS increase.
\end{abstract}
\begin{IEEEkeywords}
Fluid antenna systems, maximum ratio combining, outage probability, port correlation models, Rician fading.
\end{IEEEkeywords}
\section{Introduction}
The evolution towards sixth-generation (6G) wireless networks necessitates communication architectures capable of achieving ultra-reliable and large-scale connectivity. While massive multiple-input multiple-output (MIMO) systems have driven key advances in fifth-generation networks, pillared on massive machine-type communications, ultra-reliable low-latency communications, and enhanced mobile broadband, their reliance on fixed, uniformly spaced antenna arrays limits efficient utilization of spatial diversity \cite{10379539, 9770295}. As a result, 6G-and-beyond research is shifting towards flexible and reconfigurable array architectures that can dynamically adapt their geometry and electromagnetic characteristics to varying environments \textcolor{black}{\cite{11535420}}. Furthermore, continually increasing antenna counts becomes unsustainable due to the associated growth in power consumption and hardware complexity \cite{9349624}.

Fluid antenna systems (FASs) have recently emerged as a promising paradigm to overcome the spatial and hardware constraints of conventional massive MIMO systems \cite{9264694, 10146286, 10694739}. Typically, by enabling a single radiating element to dynamically switch among multiple predefined ports within a limited region, FASs achieve spatial diversity without requiring multiple radio-frequency (RF) chains. This makes FASs particularly suitable for compact wireless devices where deploying multiple antennas is difficult. Due to their flexibility and potential performance improvement, FASs have been recognized as an important enabling technology for 6G-and-beyond wireless systems \cite{11302793, 11247926, 11093131}.

The ports in a fluid antenna (FA) receiver are confined within a small region, with typical inter-port spacing of less than half the wavelength of the transmitted signal. The resulting spatial correlation between ports strongly influences the performance of FASs, making accurate modeling of the correlation among channel gains essential for performance analysis. For one-dimensional FA receivers, the widely used simplified channel model, adopted in \cite{9264694}, parameterizes the channels using the first physical port as the reference port. Although this model simplifies the performance analysis of FASs, it inherently imposes a restriction on the correlation structure between any two ports except the first port. \textcolor{black}{As an alternative, \cite{10103838} presented a tractable approximation of the fully correlated channel model that does not require a reference port. However, the statistics of the channel magnitudes based on this model involve nested integrals, which makes the performance analysis mathematically intractable \cite{10753482}.} To circumvent these issues, \cite{constcorr} introduced a more practical model, where a common port correlation coefficient links all ports without the need for a physical reference port, thus improving the modeling accuracy while maintaining mathematical tractability.

Such mathematical modeling of the channel gains incorporating the correlation among the ports has led to several works in recent years studying the performance of FASs. Considering the simplified channel model, \cite{9131873} studied the ergodic capacity of a FAS and presented its advantage over a traditional multi-antenna system; \cite{9833952} derived the level crossing rate, average fade duration, and ergodic capacity of a FAS; \cite{10858773} used a block-diagonal matrix formulation to study the outage probability (OP) of reconfigurable intelligent surface (RIS)-FAS; \cite{10167904} evaluated the performance of full-duplex cooperative non-orthogonal multiple access (NOMA)-based FAS; \cite{10188603} studied the OP of a FAS with port selection based on practical system delays; and \cite{9715064} proposed learning-based port selection algorithms. \textcolor{black}{On the other hand, the fully correlated channel model (or its approximations using the copula constructions) was utilized to study the performance of FASs in \cite{11459144, 10924151, 11371611, 11184548, LIU2026103261, 11646997, 11535468}}, whereas a FAS with a common port correlation coefficient was adopted in \cite{10980171, 11185052, 10855346, 10078147, 10066316}.

The majority of the above-mentioned studies, along with many others, analyze the performance of FASs assuming non-line-of-sight (NLoS) channels by modeling the channel envelopes as Rayleigh distributed. However, the presence of line-of-sight (LoS) components is an important consideration in practical systems, especially when operating between the upper mid-band and sub-THz regimes \cite{Atz25}. Thus, to account for these channel conditions, recent studies have considered Rician fading in FASs. In this regard, \cite{11098630} derived expressions for the OP of a FAS; \cite{11432102} characterized the output signal-to-noise ratio (SNR) in a continuous FAS employing matched filtering by deriving its approximate distribution in the high-SNR regime; \cite{11106811} studied the sum rate of a FAS-assisted uplink NOMA system; and \cite{11277276} analyzed the performance of a RIS-FAS. All these studies restrict their focus to the activation of a single FA port. However, the performance of FASs can be greatly enhanced by activating multiple ports through the utilization of a subset of the numerous RF chains already available in next-generation wireless terminals \cite{dash2026}. In addition to enabling multi-user communications, this capability represents a step towards realizing the concept of continuous-aperture MIMO, which requires an extremely large number of densely deployed discrete antenna elements \cite{10146286}. To this end, \cite{10375698} studied the outage performance of a FAS over Rayleigh fading channels, with a subset of the best ports selected and combined using maximum ratio combining (MRC). Similarly, \cite{10279614, 10308583} considered the addition of the channel gains at multiple ports to study the OP of a one-dimensional FAS. \textcolor{black}{The effect of selecting multiple active ports was investigated in \cite{10468625} for the secrecy outage probability of a FAS in the presence of an eavesdropper and in \cite{11558417} for the outage probability and bit error rate of a FA-assisted cooperative relaying system.} However, the impact of LoS components, prominent in ultra-dense, indoor, and vehicular 6G networks, on the performance of multi-port FASs is a crucial aspect that remains unexplored.

To address this research gap and assess the accuracy of the model with a common port correlation coefficient over the simplified channel model in the presence of a LoS component, this paper investigates a FAS where the receiver is equipped with $M$ FA ports and selects the best $K$, where the channels are subject to Rician fading. The main contributions are summarized as follows:
\begin{itemize}
\item Two different models are considered for incorporating the port correlation into the fading channel statistics: i)~the simplified channel model that treats the first physical port as the reference port, referred to in the following as {\em physical reference port model}; and ii)~the channel model that considers a common port correlation by introducing a virtual reference port, referred to in the following as {\em virtual reference port model}.
\item Considering the best $K$ among $M$ ports with the highest instantaneous SNR, combined using MRC, the OP of the FAS is derived considering both port correlation models using a characteristic function (CF) approach.
\item For both scenarios, closed-form lower bounds on the OP are obtained, from which the asymptotic OPs at high SNR are derived, revealing the diversity order of the FAS.
\item Numerical results validate the developed analytical expressions and illustrate the performance trends of the considered MRC-based FAS for different settings.
\end{itemize}

\textcolor{black}{Although \cite{10375698} also considers selecting the best $K$ of $M$ available ports, the contribution of this work lies in deriving the outage probability over Rician fading channels while explicitly accounting for the conditional channel distributions and the order statistics governing port selection. For the physical reference port model, the derivation distinguishes three selection cases according to the role of the reference port and evaluates the corresponding conditional selection and outage probabilities. These derivations also extend beyond the asymptotic characterization considered in \cite{10308583}. The analysis reveals that the performance gap between the physical and virtual reference port models widens as the number of available ports, the number of ports selected for MRC, and the average SNR increase. Comparisons with benchmark port correlation models further show that the considered models yield the same diversity order as Jakes' model, while the virtual reference port model more closely approximates the performance of the distance-dependent physical port correlation model.}

{\em Organization:} The rest of the paper is organized as follows. Section~\ref{sec:sm} details the system model of the FAS and presents the statistics of the envelopes of the channel gains based on the physical and the virtual reference port models. Section~\ref{sec:pa} presents the exact expressions for the OP of the FAS for both scenarios, derived using a CF approach. Section~\ref{sec:lbaa} presents the lower bounds and asymptotic expressions for the OP at high SNR. Section~\ref{sec:nm} details the numerical results corroborating the analytical framework, followed by the concluding remarks in Section~\ref{sec:con}.

{\em Notations:} ${\mathcal{N}} \left(\mu,\sigma^2 \right)$ and ${\mathcal{CN}} \left(\mu,\sigma^2 \right)$ denote the distributions of a real and a complex Gaussian random variable, respectively, with mean $\mu$ and variance $\sigma^2$. $\mathbb{E}\left[\cdot\right]$ denotes the statistical expectation operator, $\jmath=\sqrt{-1}$ is the imaginary unit, and $\left|\cdot \right|$ outputs the magnitude of a complex variable. ${}_1F_2(\cdot;\cdot;\cdot;\cdot)$ denotes the generalized hypergeometric function, $J_1(\cdot)$ the first-order Bessel function of the first kind, $J_0 \left( \cdot \right)$ the zero-order Bessel function of the first kind, $I_0 \left( \cdot \right)$ the zero-order modified Bessel function of the first kind, and $Q_1 \left( \cdot , \cdot \right)$ the Marcum-$Q$ function. Lastly, $\Gamma\left(\cdot,\cdot \right)$ and $\gamma\left(\cdot,\cdot \right)$ denote the upper and lower incomplete Gamma functions, respectively.
\section{System Model} \label{sec:sm}
We consider a single-antenna transmitter communicating with a receiver equipped with a one-dimensional FAS. The FAS comprises $M$ equispaced ports distributed along a linear region of length $W \lambda$, where $\lambda$ is the wavelength of the RF signal. All ports have identical marginal channel distributions and noise statistics, although their channels are spatially correlated: the following definitions therefore apply to any port indexed by $m \in \{1,\ldots,M\}$. Denoting the transmitted symbol by $s \in \mathbb{C}$, the received signal at the $m$-th port is expressed as
\beq
y_{m} = \sqrt{\alpha(d)} h_{m} s + w_{m} ,
\label{eq1}
\eeq
where $P_s = \mathbb{E} \left[ \lvert s \rvert^{2} \right]$ denotes the transmit symbol power, $\alpha(d) = \lambda^{2}/ \left(4 \pi d \right)^2$ represents the large-scale fading coefficient due to the free-space pathloss, $d$ is the distance between the transmitter and receiver, and $w_m \sim {\mathcal{CN}} \left(0,\sigma_{w}^2 \right)$ is the additive white Gaussian noise (AWGN) at the $m$-th port. Moreover, $h_m \sim {\mathcal{CN}} \left( \mu_h, \sigma_h^2 \right)$ represents the fading channel at the $m$-th port, accounting for both LoS and NLoS propagation. Thus, $\left|h_m \right|$ follows a Rician distribution with probability density function (PDF) given by
\beq
f_{|h_{m}|}(r) = \frac{2r}{\sigma_h^{2}} 
\exp \left\{-\frac{r^2 + \mu_h^{2}}{\sigma_h^{2}} \right\} 
I_{0} \left(\frac{2 \mu_h r}{\sigma_h^{2}}\right), \ r \geq 0 ,
\label{eq2}
\eeq 
with the Rician factor defined as $\kappa = \mu_h ^2/\sigma_h^2$. Without loss of generality, we consider $\mathbb{E} \left[ |h_{m}|^{2} \right]=1$, which results in the powers of the LoS and the NLoS components expressed as
\beq
\mu_h^2 = \frac{\kappa}{\kappa+1} , \quad
\sigma_h^{2} = \frac{1}{\kappa+1} ,
\label{eq3}
\eeq
respectively. Considering \eqref{eq1} along with the statistics of the channels and AWGN, the instantaneous SNR at the $m$-th port and the average SNR of the FAS are defined as
\beq
\gamma_m = \frac{\alpha (d) \left|h_m \right|^2 P_s}{\sigma_w^2} , \quad
\bar{\gamma} = \frac{\alpha (d) P_s}{\sigma_w^2} ,
\label{eq4}
\eeq
respectively, implying $\gamma_m = \bar{\gamma} \left| h_m \right|^2$.

The close spacing of the ports induces inter-port correlation, which is typically characterized using Jakes' model. Several studies on FASs adopt a simplified channel model in which the first physical port serves as the reference port \cite{9131873, 9833952, 10858773, 10167904, ganeshgc25}. In this model, the correlation coefficient between the $m$-th port and the first port is given by $J_0 \big( 2 \pi \frac{\left|m-1 \right|}{\left(M-1 \right)} W \big)$, $m \in \left\{2,\ldots,M \right\}$. Although this simplified model facilitates the analysis of FASs, it does not reproduce the distance-dependent correlation between arbitrary pairs of non-reference ports. Specifically, Jakes' model prescribes a correlation coefficient $J_0 \big( 2 \pi \frac{\left|m-\ell \right|}{\left(M-1 \right)} W \big)$ between the $m$-th and and $\ell$-th ports, whereas the simplified model results in $J_0 \big( 2 \pi \frac{\left|m-1 \right|}{\left(M-1 \right)} W \big) J_0 \big( 2 \pi \frac{\left|\ell-1 \right|}{\left(M-1 \right)} W \big)$. Thus, the correlation between two non-reference ports depends on their respective distances from the reference port rather than solely on their mutual separation. As an alternative approximation, we adopt a common port correlation coefficient \cite{constcorr}, given by
\beq
\rho = \sqrt{2 \left( {}_1F_2 \left(\frac{1}{2}; 1;
\frac{3}{2}; -\pi^2 W^2\right)
- \frac{J_1(2\pi W)}{2\pi W} \right)} .
\label{eq5}
\eeq
Although \eqref{eq5} captures the correlation among all ports without the need for a physical reference port, incorporating it to reflect the correlation between channels can be achieved by considering a virtual reference port. The following subsections describe the channel statistics of the FAS using both the {\em virtual reference port model}, which adopts a common port correlation coefficient, and, for comparison, the widely used {\em physical reference port model}, which considers the first physical port as the reference port.
\subsection{Channel Statistics with the Virtual Reference Port Model}
Considering the virtual reference port model, we introduce a reference port whose channel $h_0$ follows a zero-mean complex Gaussian distribution, i.e., $h_0 \sim \mathcal{CN} \left( 0, \sigma_h^2 \right)$. Thus, the channels can be expressed in terms of $h_0$ as
\beq
{\color{black}{h_m = \mu_h + \rho h_0 + \sqrt{1-\rho^2} c_m , 
\ m \in \{1, \ldots, M\} ,}}
\label{eq6}
\eeq
where $\{c_m\}_{m=1}^{M}$ are independent $\mathcal{CN}(0, \sigma_h^2)$ random variables. Furthermore, from \eqref{eq6}, we have that $h_m | h_0$ follows a non-zero-mean complex Gaussian distribution, i.e.,
\beq
h_m | h_0 \sim \mathcal{CN}
\left(\mu_h + {\color{black} \rho} h_0, \left(1- {\color{black} \rho^2} \right) \sigma_h^2 \right) .
\label{eq9}
\eeq
\textcolor{black}{Let us define}
\beqarr
{\color{black} Y_0
= \bar{\gamma} \left| \mu_h + \rho h_0 \right|^2 ,}
\label{eq10}
\eeqarr
\textcolor{black}{whose PDF is obtained from the statistics of $h_0$ as
\beq
f_{Y_0} \left(y_0 \right)
= \zeta_0 \, e^{-\zeta_0 \left(y_0+\delta \right)}
I_0 \! \big(2\zeta_0 \sqrt{\delta y_0} \big) , \ y_0 \ge 0 ,
\label{eq8}
\eeq
with $\delta=\bar\gamma\mu_h^2$ and $\zeta_0 = (\kappa+1)/(\bar\gamma\rho^2)$. Conditioned on $Y_0=y_0$, $\gamma_m$ follows a non-central chi-squared distribution with two degrees of freedom and non-centrality $y_0$, resulting in the PDF
\beq
\hspace{-2mm} f_{\gamma_m \mid Y_0 = y_0}(v_m)
= \zeta \, e^{-\zeta \left(v_m + y_0 \right)}
I_0 \! \left( 2 \zeta \sqrt{v_m y_0} \right) , \ v_m \ge 0 ,
\label{eq11}
\eeq
with $\zeta = (\kappa+1)/\left(\bar\gamma(1-\rho^2) \right)$. Using the series-form expansion $I_0(x) = \sum_{n=0}^{\infty} (x/2)^{2n}/(n!)^2$, we can rewrite \eqref{eq11} as
\beq
f_{\gamma_m | Y_0 = y_0} \left( v_m \right)
= \sum_{n=0}^{\infty} d_n \left( y_0 \right) v_m^n e^{-\zeta v_m} ,
\label{eq12}
\eeq
with
\beq
d_n \left(y_0 \right)
= \frac{\zeta^{2n+1} \, y_0^n \, e^{-\zeta y_0}}{(n!)^2} .
\label{eq13}
\eeq
Additionally, using \eqref{eq12}, the cumulative distribution function (CDF) of $\gamma_m | Y_0$ is obtained as
\beq
F_{\gamma_m \mid Y_0 = y_0} \left(v_m \right)
= \sum_{n=0}^{\infty} d_n(y_0) \, \frac{\gamma \left(n+1,\zeta v_m \right)}{\zeta^{n+1}} .
\label{eq14}
\eeq}
\subsection{Channel Statistics with the Physical Reference Port Model}
Considering the physical reference port model, the channels can be expressed using their distributions as
\begin{align}
h_{1} & = \sigma_h x_{1} + \jmath \sigma_h y_{1} + \mu_h , \nn \\
h_{m} & = \sigma_h \big( \sqrt{1-\rho^{2}}\,x_{m} + {\color{black}{\rho}} x_{1} \big) + \jmath \sigma_h \big( \sqrt{1-\rho ^{2}}\,y_{m} + {\color{black}{\rho}} y_{1} \big) \nn \\
& \quad + \mu_h , \ m \in \{ 2,\ldots,M\} ,
\label{eq15}
\end{align}
where $\{x_{m}, y_{m}\}_{m=1}^{M}$ are independent $\mathcal{N} \left( 0, 1/2 \right)$ random variables. From \eqref{eq15}, we observe that $\left| h_1 \right|^2$ follows a non-central chi-squared distribution with two degrees of freedom and non-centrality $\mu_h^2$. This results in the PDF of $\left| h_1 \right|^2$ obtained as
\beq
f_{|h_1|^2} \left(u \right) = \frac{1}{\sigma_h^2}
\exp \left\{ \! - \frac{u+ \mu_h^2}{\sigma_h^2} \! \right\}
\! I_0 \! \left( \frac{2\mu_h\sqrt{u}}{\sigma_h^2} \right) , \ u \ge 0 .
\label{eq16}
\eeq
From \eqref{eq4}, the instantaneous SNR at the first port is $\gamma_1 = \bar{\gamma} |h_1|^2$. Using this relation along with \eqref{eq16} results in the PDF of $\gamma_1$ given by
\begin{align}
f_{\gamma_1} \left( v_1 \right)
& = \frac{1}{\bar{\gamma}}
f_{|h_1|^2}\!\left(\frac{v_1}{\bar{\gamma}}\right)
= \frac{\kappa+1}{\bar{\gamma}}
\exp \left\{ -\kappa-\frac{\kappa+1}{\bar{\gamma}} v_1 \right\} \nn \\
&\quad \times
I_{0} \left( 2\sqrt{\frac{\kappa \left(\kappa+1 \right)}
{\bar{\gamma}} v_1} \right) , \, \ v_1 \ge 0 .
\label{eq17}
\end{align}
From \eqref{eq15}, we further have that $h_m | h_1$, $m \in \left\{2,\ldots,M \right\}$, follows a non-zero-mean complex Gaussian distribution, i.e.,
\beq
h_m | h_1 \sim \mathcal{CN} \left( \mu_h \left(1-\rho \right) + \rho h_1 , \left(1-{\color{black} \rho^2} \right) \sigma_h^2 \right) .
\label{eq18}
\eeq
{\color{black}{Writing $h_1 = \sqrt{v_1/\bar{\gamma}} \, e^{\jmath \theta_1}$, the standard bivariate-Gaussian-in-polar decomposition of a Rician-distributed complex variable gives
\beq
f \left(\theta_1 | \gamma_1 = v_1 \right)
= \frac{e^{\xi (v_1) \cos \theta_1}}{2 \pi I_0 \left( \xi (v_1) \right)} , \
 \xi (v_1) = 2\zeta \left(1-\rho^2 \right) \sqrt{\delta v_1} .
\label{eqC1}
\eeq
This results in the noncentrality parameter of $\gamma_m$, $m \in \{2,\ldots, M \}$, conditioned on $h_1$ given by
\beq
\Delta \left(v_1,\theta_1 \right)
\! = \! \bar{\gamma} \left| \rho h_1 + \mu_h \left(1-\rho \right) \right|^2
\! = A (v_1) + B (v_1) \cos \theta_1 ,
\label{eqC2}
\eeq
with $A (v_1) = \rho^2 v_1 + \delta (1-\rho)^2$ and $B (v_1) = 2 \rho (1-\rho) \sqrt{\delta v_1}$. Conditioned on $(v_1,\theta_1)$ jointly, $\{\gamma_m\}_{m=2}^M$ are i.i.d. non-central chi-squared random variables, resulting in the PDF
\beq
f_{\gamma_m | v_1,\theta_1} \left(v_m \right)
= \zeta e^{-\zeta \left(v_m + \Delta (v_1,\theta_1) \right)}
I_0 \! \big(2\zeta \sqrt{v_m \Delta (v_1,\theta_1)} \big) .
\label{eq19}
\eeq
Using the same series-form expansion as in \eqref{eq12}, the PDF of $\gamma_m | v_1, \theta_1$, $m \in \{2,\ldots,M\}$, can be alternatively expressed as
\beq
f_{\gamma_m | v_1,\theta_1} \left(v_m \right)
= \sum_{n=0}^{\infty} d_{n} (v_1,\theta_1) v_m^{n} e^{-\zeta v_m} , \
v_m \geq 0 ,
\label{eq21}
\eeq
with
\beq
d_{n} (v_1,\theta_1) = \frac{\zeta^{2n+1} \Delta^n(v_1,\theta_1) \, e^{-\zeta \Delta (v_1,\theta_1)}}{(n!)^2} .
\label{eq22}
\eeq
Lastly, using \eqref{eq21}, the CDF of $\gamma_m | v_1,\theta_1$, $m \in \{2,\ldots,M\}$, is obtained as 
\beq
F_{\gamma_{m} | v_1,\theta_1} \! \! \left( v_m \right)
= \sum_{n=0}^{\infty} d_n(v_1,\theta_1) \, \frac{\gamma \left(n+1,\zeta v_m \right)}{\zeta^{n+1}} .
\label{eq23}
\eeq
}}
\section{Performance Analysis} \label{sec:pa}
Upon receiving the data symbol, the receiver activates the best $K \in \{1, \ldots, M\}$ ports with the highest instantaneous SNR among the $M$ available FA ports. In practice, the receiver can observe the SNR at each port via analog power/envelope measurement prior to the RF chains. Moreover, activating $K$ ports requires $K$ RF chains, which improves the performance at the cost of increased hardware complexity of the FAS.\footnote{\textcolor{black}{The receiver uses a switch network to connect the $M$ ports to $K$ RF chains, followed by digital MRC. This hybrid selection/MRC architecture, also known as generalized selection combining, requires fewer RF chains than a traditional array with $M$ fixed antennas, and setting $K=1$ recovers the single-RF-chain FAS. During the preamble, the SNRs at the ports are estimated by sequentially connecting a measurement path (implemented using a power detector or one RF chain) to each port. This procedure requires $M$ measurement slots and the associated switching time. In this regard, PIN-diode switches support microsecond-level switching \cite{10740058}, while measuring only a subset of ports can further reduce the overhead \cite{dinis2026}. We assume that the resulting SNR ranking remains valid over the whole data block and neglect channel state information (CSI) aging, switching losses, mutual coupling, and calibration errors, so the resulting performance serves as benchmarks under ideal port selection \cite{10375698}.}} Since the port selection depends on the relative ordering of the SNR values, we denote the SNRs arranged in descending order as $\gamma_{[1]} > \gamma_{[2]} > \ldots > \gamma_{[M]}$, where $\gamma_{[m]}$ represents the $m$-th largest instantaneous SNR among the $M$ ports. Based on this ordering, the indices corresponding to the selected ports are grouped into the set $\mathcal{S} = \left\{ [1], [2], \ldots, [K] \right\}$ representing the best $K$ ports, while the indices corresponding to the remaining unselected ports are grouped into the set $\bar{\mathcal{S}} = \left\{ [K+2], [K+3], \ldots, [M] \right\}$. In this setting, $\nu = \gamma_{[K+1]}$ represents the largest instantaneous SNR among the unselected ports, which defines the boundary between the selected and remaining ports. The signals from the $K$ selected ports are then combined using MRC, with post-combining SNR given~by 
\beq
\gamma_{\text{MRC}} = \sum_{m=1}^{K} \gamma_{[m]} 
= \bar{\gamma} \sum_{m=1}^{K} \left| h_{[m]} \right|^{2} .
\label{eq24}
\eeq
Thus, the OP of the MRC-based FAS can be expressed as
\beq
P_{\text{out}}
= \Pr \left( \gamma_{\text{MRC}} \leq \gamma_{\text{th}} \right)
= \Pr \left( \sum_{m=1}^{K} \gamma_{[m]} \leq \gamma_{\text{th}} \right) .
\label{eq25}
\eeq
We derive the OP for the FAS with the virtual and physical reference port models in Sections~\ref{sec:OP_virtual} and~\ref{sec:OP_physical}, respectively.
\subsection{OP Analysis with the Virtual Reference Port Model} \label{sec:OP_virtual}
Considering the virtual reference port model, the OP can be defined from \eqref{eq25} as
\beq
\color{black}{P_{\text{out}}^{\text{v.r.p.}}
= \mathbb{E}_{Y_0} \left[ \Pr \left( \left.
\sum_{m \in \mathcal{S}} \gamma_m \le \gamma_{\text{th}}
\right| Y_0 = y_0 \right) \right] ,}
\label{eq26}
\eeq
where the conditioning on $Y_0$ in \eqref{eq26} is considered to simplify the analysis by exploiting the independence among $\{\gamma_m | Y_0\}_{m=1}^{M}$. Thus, the conditional probability of the instantaneous SNR of the remaining $U$ ports to be smaller than $\nu$ conditioned on $Y_0$ can be expressed as
\beq
{\mathcal{G}} \left( \nu, y_0 \right)
= \Pr \left( \gamma_{m} \leq \nu , \, m \in \bar{\mathcal{S}}
| \, Y_{0} = y_0 \right) .
\label{eq27}
\eeq
Furthermore, the conditional probability of the best $K$ ports being selected and the event of an outage occurring conditioned on $Y_0$ is obtained as
\beqarr
{\mathcal{H}} \left(\gamma_{\text{th}}, \nu, y_0 \right)
\! = \Pr \left( \! \left. \sum_{m \in \mathcal{S}} \! \gamma_{m} \leq \gamma_{\text{th}} ,
\gamma_{m} > \nu \right| Y_{0} = y_0 \right) .
\label{eq28}
\eeqarr

From the statistics of $\{h_m\}_{m=1}^{M}$, we note that the channels are exchangeable random variables. This implies that there exist ${{M}\choose{K}}$ combinations for the selected ports. Consequently, there would be $U+1$ ports available for the choice with the $(K+1)$-th highest channel gain. This allows the expression of the OP in \eqref{eq26} to be rewritten as
\begin{align}
P_{\text{out}}^{\text{v.r.p.}}
& = \mathbb{E}_{Y_0} \bigg[ \int_{0}^{\infty} 
{{M}\choose{K}} \left(U+1 \right) 
{\mathcal{H}} \left( \gamma_{\text{th}} , \nu, y_0 \right) \nn \\
& \quad
\times {\mathcal{G}} \left( \nu, y_0 \right)
f_{ \nu | Y_0 = y_0} (\nu) \text{d} \nu \bigg] .
\label{eq29}
\end{align}
For the scenario where all $M$ ports are selected, the OP simplifies to $P_{\text{out} | K=M}^{\text{v.r.p.}} \left( \gamma_{\text{th}} \right) = \mathbb{E}_{Y_0} \left[ {\mathcal{H}} \left( \gamma_{\text{th}},0,y_0 \right) \right]$. Thus, to compute the OP in \eqref{eq29}, we next derive the expressions in \eqref{eq27} and \eqref{eq28}.

Owing to the statistical independence of $\{\gamma_m | Y_0\}_{m=1}^{M}$ and using \eqref{eq10}, \textcolor{black}{the joint PDF of $\{\gamma_m\}_{m \in \bar{\mathcal{S}}}$ conditioned on $Y_0$ is given by $\prod_{m=1}^{U} f_{\gamma_m | Y_0=y_0}(v_m)$,} which yields
\begin{align}
& \! {\mathcal{G}} \left( \nu, y_0 \right) \nn \\
& = \int_{0}^{\nu} \! \! \! \ldots \! \int_{0}^{\nu}
\! \! \! f_{\gamma_{[K+2]},\ldots, \gamma_{[M]} | Y_{0}}
\! \! \left( v_{K+2},\ldots,v_{M} \right)
\text{d} v_{K+2} \ldots \text{d} v_M \nn \\
& = {\color{black} \prod_{m=1}^U \!  F_{\gamma_m \mid Y_0 = y_0} \left(v \right) = F^U (\nu, y_0)
= F^{M-K-1} (\nu, y_0)}. \!
\label{eq31}
\end{align}
To derive the expression of ${\mathcal{H}} \left( \gamma_{\text{th}}, \nu, y_0 \right)$, we need the PDF of the random variable involved in \eqref{eq29}. To this end, we first obtain the CF of $\gamma_m | Y_0$ under the port activation constraint $\gamma_m > \nu$, which can be obtained using \eqref{eq12} as
\begin{align}
\! \! \! \! \Psi_{\gamma_m | Y_0}
\left( \jmath \omega ; \nu, y_0 \right)
& = \mathbb{E} \left[ \left. e^{\jmath \omega \gamma_m} \right|  
Y_0 , \gamma_m > \nu \right] \nn \\
& = \int_{\nu}^{\infty} e^{\jmath \omega v_m}
f_{\gamma_m | Y_0 = v_0} \left( v_m \right)
\text{d} v_m \nn \\
& = \sum_{n=0}^{\infty} d_n \left( y_0 \right)
\int_{\nu}^{\infty} v_m^n
e^{- \left( \zeta - \jmath \omega \right) v_m} \text{d} v_m \nn \\
& = \sum_{n=0}^{\infty} d_n \left( y_0 \right)
\frac{\Gamma \left(n+1, \left( \zeta - \jmath \omega \right) \nu \right)}
{\left( \zeta - \jmath \omega \right)^{n+1}} .
\label{eq32}
\end{align}
Using the series expansion $\Gamma \left(n+1,z \right) = n! e^{-z} \sum_{\ell=0}^{n} \frac{z^{\ell}}{\ell!}$, the expression of $\Psi_{\gamma_m | Y_0}
\left( \jmath \omega ; \nu, y_0 \right)$ can be rewritten as 
\begin{align}
& \! \! \! \! \Psi_{\gamma_m | Y_0}
\left(\jmath \omega ; \nu, y_0 \right) \nn \\
& \ = e^{-\zeta \nu} e^{\jmath \omega \nu}
\sum_{n=0}^{\infty} d_n \left( v_0 \right) n!
\sum_{\ell=0}^{n} \frac{\nu^{\ell}}
{\ell! \left( \zeta - \jmath \omega \right)^{\left(n+1-\ell \right)}} .
\label{eq35}
\end{align}
Since the selected ports are conditionally independent given $Y_0$, the CF of the sum of the instantaneous SNRs for the $K$ selected ports, conditioned on $Y_0$ and the constraint $\gamma_m > \nu$, is obtained using \eqref{eq35} as
\begin{align}
& \! \! \! \! \! \! \! \!
\Psi_{\sum\limits_{m \in {\mathcal{S}}} \gamma_m | Y_0}
\left( \jmath \omega ; \nu, y_0 \right) \nn \\
& = \prod_{m=1}^{K} \Psi_{\gamma_m | Y_0}
\left( \jmath \omega ; \nu, y_0 \right)
= \Psi_{\gamma_m | \gamma_0}^K
\left( \jmath \omega ; \nu, v_0 \right) \nn \\
& = e^{-\zeta K \nu} e^{\jmath \omega K \nu}
\sum_{n_1=0}^{\infty} \ldots \sum_{n_K=0}^{\infty}
\left( \prod_{m=1}^{K} d_{n_m} \left( y_0 \right) n_m! \right) \nn \\
& \quad \times \sum_{\ell_1=0}^{n_1} \ldots \sum_{\ell_K=0}^{n_K}
\left( \prod_{m=1}^{K} \frac{\nu^{\ell_m}}{\ell_m!} \right)
\left( \zeta - \jmath \omega \right)^{-\eta_K \left( \mathbf{n}, \boldsymbol{\ell} \right)} ,
\label{eq36}
\end{align}
with $\mathbf{n} = [n_1,\ldots,n_K]^T$, $\boldsymbol{\ell} = [\ell_1,\ldots, \ell_K]^T$, and
\beq
\eta_K \left( \mathbf{n}, \boldsymbol{\ell} \right)
= K + \sum_{m=1}^{K} n_m - \sum_{m=1}^{K} \ell_m .
\label{eq38}
\eeq
The PDF of $\sum_{m \in {\mathcal{S}}} \gamma_m | Y_0$ with $\{\gamma_m > \nu\}_{m \in \mathcal{S}}$ can thus be computed from \eqref{eq38} as
\begin{align}
f_{\sum\limits_{m \in \mathcal{S}} \gamma_m ,
\gamma_m > \nu | Y_0} \left( u \right)
& = \int_{-\infty}^{\infty}
\! \! \! e^{- \jmath \omega u} \,
\Psi_{\sum\limits_{m \in {\mathcal{S}}} \gamma_m | Y_0}
\left( \jmath \omega ; \nu, y_0 \right) \text{d} \omega \nn \\
& = e^{-\zeta K \nu} \! \!
\sum_{n_1=0}^{\infty} \! \! \ldots \! \! \! \sum_{n_K=0}^{\infty}
\!\!\!\left( \prod_{m=1}^{K} d_{n_m}(y_0) n_m! \! \right) \nn \\
& \quad \times \sum_{\ell_1=0}^{n_1} \ldots \sum_{\ell_K=0}^{n_K}
\left( \prod_{m=1}^{K} \frac{\nu^{\ell_m}}{\ell_m!} \right) \nn \\
& \quad \times \int_{-\infty}^{\infty}
\frac{e^{- \jmath \omega \left( u- K\nu \right)}}
{\left( \zeta - \jmath \omega \right)^{\eta_K \left( \mathbf{n}, \boldsymbol{\ell} \right)}} \text{d} \omega, 
\label{eq40}
\end{align}
and evaluating the integral results allows to express \eqref{eq40} as
\begin{align}
& \! \! \! f_{\sum\limits_{m \in \mathcal{S}} \gamma_m ,
\gamma_m > \nu | Y_0} \left(u \right)
= e^{-\zeta K \nu} \!
\sum_{n_1=0}^{\infty} \! \! \ldots \! \! \! \sum_{n_K=0}^{\infty} \! \! \!
\left( \prod_{m=1}^{K} \! d_{n_m} (y_0)  k_m! \! \right) \nn \\
& \! \! \! \times \sum_{\ell_1=0}^{n_1} \ldots \sum_{\ell_K=0}^{n_K}
\left( \prod_{m=1}^{K} \frac{\nu^{\ell_m}}{\ell_m!} \right)
\frac{\left(u - K \nu \right)^{\eta_K \left( \mathbf{n}, \boldsymbol{\ell} \right)-1}}
{\left( \eta_K \left( \mathbf{n}, \boldsymbol{\ell} \right) - 1 \right)!}
e^{-\zeta \left( u-K \nu \right)} , \nn \\
& \hspace{6.5cm} u > K \nu .
\label{eq41}
\end{align}
Using \eqref{eq41}, the expression of $ {\mathcal{H}} \left( \gamma_{\text{th}}, \nu, y_0 \right)$ in \eqref{eq28} can be written as
\begin{align}
{\mathcal{H}} \left( \gamma_{\text{th}}, \nu, y_0 \right)
& = \int_{K \nu}^{\gamma_{\text{th}}}
f_{\sum\limits_{m \in \mathcal{S}} \gamma_m ,
\gamma_m > \nu | Y_0} \left(u \right) \text{d} u \nn \\
& = e^{-\zeta K \nu}
\sum_{n_1=0}^{\infty} \ldots \sum_{n_K=0}^{\infty}
\left( \prod_{m=1}^{K} d_{n_m} \left(y_0 \right) k_m! \right) \nn \\
& \quad \times \sum_{\ell_1=0}^{n_1} \ldots \sum_{\ell_K=0}^{n_K}
\left( \prod_{m=1}^{K} \frac{\nu^{\ell_m}}{\ell_m!} \right) \nn \\
&\quad \times \frac{ \gamma \big(
\eta_K \left( \mathbf{n},\boldsymbol{\ell} \right),
\zeta ( \gamma_{\text{th}} - K \nu )\big)}
{\big( \eta_K \left( \mathbf{n}, \boldsymbol{\ell} \right) - 1 \big)! \, 
\zeta^{\eta_K \left( \mathbf{n},\boldsymbol{\ell} \right)}} ,
\label{eq45}
\end{align}
with ${\mathcal{H}} \left( \gamma_{\text{th}}, \nu, y_0 \right)=0$ for $\gamma_{\text{th}} < K \nu$. Finally, substituting \eqref{eq31} and \eqref{eq45} into \eqref{eq29} results in the expression of the OP given by
\begin{align}
P_{\text{out}}^{\text{v.r.p.}} \left(\gamma_{\text{th}} \right)
& = \int_{0}^{\infty} \int_{0}^{\gamma_{\text{th}}/K}
\binom{M}{K} \left( M-K \right) e^{-\zeta K \nu} \nn \\
&\quad \times
\sum_{n_1=0}^{\infty} \ldots \sum_{n_K=0}^{\infty}
\left(\prod_{m=1}^{K} d_{n_m} \left( y_0 \right) k_m! \right) \nn \\
& \quad \times \sum_{\ell_1=0}^{n_1} \ldots \sum_{\ell_K=0}^{n_K}
\left( \prod_{m=1}^{K} \frac{\nu^{\ell_m}}{\ell_m!} \right) \nn \\
& \quad \times \frac{\gamma \big(
\eta_K \left(\mathbf{n},\boldsymbol{\ell} \right),
\zeta ( \gamma_{\text{th}} - K \nu ) \big)}
{\big(\eta_K \left(\mathbf{n},\boldsymbol{\ell} \right)-1 \big)! \,
\zeta^{\eta_K \left(\mathbf{n},\boldsymbol{\ell} \right)}}
{\color{black} F^{M-K-1} (\nu,y_0)} \nn \\
& \quad \times f_{\nu | Y_0 = y_0} \left( \nu \right)
f_{Y_0} \left( y_0 \right) \text{d} \nu \, \text{d} y_0 .
\label{eq46}
\end{align}

{\em Case of $K=M$:} When the FA receiver selects all ports, the OP in \eqref{eq29} reduces to 
\beq
P_{\text{out} | K=M}^{\text{v.r.p.}}
= \int_{0}^{\infty} {\mathcal{H}}
\left( \gamma_{\text{th}}, 0, y_0 \right)
f_{Y_0} \left( y_0 \right) \text{d} y_0 .
\label{eq47}
\eeq
For $\nu=0$ and $K=M$, $\{\ell_m\}_{m=1}^{K}$ in \eqref{eq38} are all equal to zero and the expression of $\eta_M \left( \undb{n} , \boldsymbol{\ell} \right)$ simplifies to $M + \sum_{m=1}^M n_m$. This allows to simplify \eqref{eq45} as
\begin{align}
& \! \! \! {\mathcal{H}} \left( \gamma_{\text{th}}, 0, y_0 \right)
= \sum_{n_1=0}^{\infty} \ldots \sum_{n_M=0}^{\infty}
\left( \prod_{m=1}^{K} d_{n_m} \left( y_0 \right) n_m! \right) \nn \\
& \qquad \qquad \times \frac{\gamma \big(M + \sum_{m=1}^{M} n_m ,
\zeta \gamma_{\text{th}} \big)}
{ \big(M + \sum_{m=1}^{M} n_m - 1 \big)! \,
\zeta^{M + \sum_{m=1}^{M} n_m}} .
\label{eq54}
\end{align}
By further substituting \eqref{eq54} and \eqref{eq8} into \eqref{eq47} and integrating with respect to $y_0$ by using the same series-form expansion as in \eqref{eq12}, the final OP expression with $K=M$ for the FAS with the virtual reference port model is obtained as
{\color{black}{
\begin{align}
P_{\text{out} | K=M}^{\text{v.r.p.}}
& = e^{- \delta \zeta_0}
\sum_{n_1=0}^{\infty} \ldots \sum_{n_M=0}^{\infty}
\frac{\gamma \left(M + \sum\limits_{m=1}^{M} n_m ,
\zeta \gamma_{\text{th}} \right)}
{\left( M \! + \! \! \sum\limits_{m=1}^{M} n_m - 1 \right)!} \nn \\
& \hspace{-7mm} \times \frac{\zeta^{\sum\limits_{m=1}^M n_m}}{\left( \prod\limits_{m=1}^M n_m! \right)}
\sum_{k=0}^{\infty}
\frac{\zeta_0^{2k+1} \delta^k
\left(k + \sum\limits_{m=1}^M n_m \right)!}
{(k!)^2 \left(\zeta M + \zeta_0 \right)^{k + \sum\limits_{m=1}^M n_m + 1}} .
\label{eq55}
\end{align}
Finally, utilizing the relation $\sum_{n_1+\ldots+n_M=n} \frac{1}{\prod_{m=1}^M n_m!} = \frac{M^n}{n!}$, the expression in \eqref{eq55} can be further simplified as
\begin{align}
P_{\text{out} | K=M}^{\text{v.r.p.}}
& = e^{- \delta \zeta_0}
\sum_{n=0}^{\infty}
\frac{\left( M \zeta \right)^n
\gamma \left(M + n , \zeta \gamma_{\text{th}} \right)}
{\left( M + n - 1 \right)! n!} \nn \\
& \quad \times \sum_{k=0}^{\infty}
\frac{\zeta_0^{2k+1} \delta^k
\left(k + n \right)!}
{(k!)^2 \left(M \zeta + \zeta_0 \right)^{k + n + 1}} .
\label{eq55new}
\end{align}
}}
\subsection{OP Analysis with the Physical Reference Port Model} \label{sec:OP_physical}
Considering the physical reference port model, the OP can be defined from \eqref{eq25} as
\textcolor{black}{
\beq
P_{\text{out}}^{\text{p.r.p.}}
= \mathbb{E}_{\gamma_1} \left[ \mathbb{E}_{\theta_1 \mid \gamma_1}
\left[ \Pr \left( \left. \sum_{m \in \mathcal{S}} \gamma_m
\le \gamma_{\text{th}} \right| \gamma_1 = v_1 , \theta_1 \right) \right] \right] ,
\label{eq56}
\eeq
where the conditioning on $(v_1, \theta_1)$ in \eqref{eq56} is considered to simplify the analysis by exploiting the independence among $\{\gamma_m | v_1, \theta_1 \}_{m=2}^{M}$. For brevity, we write $F \left( \nu, v_1, \theta_1 \right) = F_{\gamma_m \mid v_1, \theta_1 }(\nu)$ for the CDF of the SNR of a non-reference port given in \eqref{eq23}.} 

{\color{black}{With port~$1$ representing the reference port, its SNR $\gamma_1$ is itself one of the $M$ candidates and may rank among the best $K$ ports. Conditioned on $(v_1,\theta_1)$, the remaining SNRs $\{\gamma_m\}_{m=2}^{M}$ are i.i.d. and mutually exchangeable by \eqref{eq21}. Let $\nu = \gamma_{[K+1]}$ denote the largest unselected SNR. Define $U' = M-K-2$ and $U''=M-K-1$, with $U \in \{U',U''\}$. The rank of port~$1$ relative to this boundary port gives rise to three mutually exclusive and exhaustive cases:
\begin{itemize}
\item[\textcolor{black}{(a)}] \underline{Case of $\nu > v_1$}: Port~$1$ is unselected and is not the boundary port, implying that the boundary port and the $K$ selected ports belong to the $M-1$ non-reference ports. Choosing which one of these $M-1$ ports is the boundary, and which $K$ of the remaining $M-2$ rank above it, gives $(M-1) {{M-2} \choose {K}} = {{M-1} \choose {K}} (U'+1)$ combinations.
\item[\textcolor{black}{(b)}] \underline{Case of $\nu < v_1$}: Port~$1$ is selected and occupies one of the $K$ selected positions. The remaining $K-1$ selected ports, the boundary port, and the $U''$ unselected ports belong to the $M-1$ non-reference ports, giving $(M-1) {{M-2} \choose {K-1}} = {{M-1} \choose {K-1}} (U''+1)$ combinations.
\item[\textcolor{black}{(c)}] \underline{Case of $\nu = v_1$}: Port~$1$ is the boundary port and is therefore unselected. Exactly $K$ of the $M-1$ non-reference ports exceed $v_1$ and form the selected set, while the remaining $M-K-1$ fall below $v_1$. The selected set can be chosen in ${{M-1} \choose {K}}$ ways.
\end{itemize}
Accordingly, we define the conditional outage probabilities for $K$ and $K-1$ i.i.d. non-reference ports whose SNRs exceed $\nu$ as
\bsub
\begin{align}
{\mathcal{H}}_{\textrm{a}} \left(\gamma_{\text{th}}, \nu, v_1,\theta_1 \right)
= \Pr \left( \left. \sum_{m=1}^{K} \gamma_{m} \leq \gamma_{\text{th}} ,
\gamma_{m} > \nu \right| v_1,\theta_1 \right) ,
\label{eq58} \\
{\mathcal{H}}_{\textrm{b}} \left(\gamma_{\text{th}}, \nu, v_1,\theta_1 \right)
= \Pr \left( \left. \sum_{m=1}^{K-1} \gamma_{m} \leq \gamma_{\text{th}} ,
\gamma_{m} > \nu \right| v_1,\theta_1 \right) ,
\label{eqD1}
\end{align}
\esub
where the $K$ (respectively $K-1$) ports in \eqref{eq58} (respectively \eqref{eqD1}) are i.i.d.\ draws from \eqref{eq21}, with ${\mathcal{H}}_{\textrm{b}}(\gamma_{\text{th}},\nu,v_1,\theta_1) = 1$ for $K=1$. This results in the conditional OP expression in \eqref{eq56} to be rewritten as
\bsub
\begin{align}
\Pr \! \left( \! \left. \sum_{m \in \mathcal{S}}
\! \gamma_m \! \le \! \gamma_{\text{th}} \right| v_1, \theta_1 \! \right)
& = P_{\textrm{a}} \left( \gamma_{\text{th}}, v_1,\theta_1 \right)
+ P_{\textrm{b}} \left( \gamma_{\text{th}}, v_1,\theta_1 \right) \nn \\
& \quad + P_{\textrm{c}} \left( \gamma_{\text{th}}, v_1,\theta_1 \right) ,
\label{eq59}
\end{align}
with
\begin{align}
P_{\textrm{a}} \left(\gamma_{\text{th}}, v_1,\theta_1 \right)
& = {{M-1} \choose {K}} \left(U'+1 \right)
\int_{v_1}^{\gamma_{\text{th}}/K}
F^{U'}(\nu,v_1,\theta_1) \nn \\
& \quad
\times {\mathcal{H}}_{\textrm{a}} \left(\gamma_{\text{th}},\nu,v_1,\theta_1 \right)
f_{\gamma_m | v_1,\theta_1}(\nu) \text{d} \nu ,
\label{eqD2}
\end{align}
\begin{align}
P_{\textrm{b}} \left(\gamma_{\text{th}}, v_1,\theta_1 \right)
& = {{M-1} \choose {K-1}} \left(U''+1 \right)
\int_{0}^{v_1} F^{U''}(\nu,v_1,\theta_1) \nn \\
& \quad \times {\mathcal{H}}_{\textrm{b}} \left(\gamma_{\text{th}}-v_1,\nu,v_1,\theta_1 \right)
f_{\gamma_m | v_1,\theta_1}(\nu) \text{d} \nu ,
\label{eqD3} \\
P_{\textrm{c}} \left(\gamma_{\text{th}}, v_1,\theta_1 \right)
& = {{M-1} \choose {K}} F^{M-1-K}(v_1,v_1,\theta_1) \nn \\
& \quad
\times {\mathcal{H}}_{\textrm{a}} \left(\gamma_{\text{th}}, v_1, v_1,\theta_1 \right) ,
\label{eqD4}
\end{align}
\esub
where the integral in \eqref{eqD3} is implicitly restricted to $\nu \le (\gamma_{\text{th}}-v_1)/(K-1)$ for $K>1$ (since we have ${\mathcal{H}}_{\textrm{b}}=0$ outside this range). For $K=M$, all ports are combined, and there is no unselected port. Cases (a) and (c) therefore cannot occur, as they require port~$1$ to be unselected or to be the boundary port, implying $P_{\textrm{a}}=P_{\textrm{c}}=0$. Only case (b) remains, in which port~$1$ is always selected, resulting in $P_{\text{out} | K=M}^{\text{p.r.p.}} \left( \gamma_{\text{th}} \right) = \mathbb{E}_{\gamma_1} \left[ \mathbb{E}_{\theta_1 \mid \gamma_1} \left[ {\mathcal{H}}_{\textrm{b}} \left( \gamma_{\text{th}}-v_1, 0, v_1,\theta_1 \right) \right] \right]$.

We next derive the expressions in \eqref{eqD2} and \eqref{eqD3}, for which we need the PDF of the involved random variable. Owing to the statistical independence of $\{\gamma_m | v_1,\theta_1\}_{m=2}^{M}$ and using \eqref{eq21}, the joint PDF of any $J$ non-reference ports conditioned on $(v_1,\theta_1)$ is given by $\prod_{j=1}^{J} f_{\gamma_m | v_1,\theta_1}(v_j)$.}}
To this end, we first obtain the CF of $\gamma_m | v_1,\theta_1$ under the port activation constraint $\gamma_m > \nu$, which can be obtained using \eqref{eq19} as
\begin{align}
\Psi_{\gamma_m | v_1,\theta_1} &
\left( \jmath \omega ; \nu, v_1,\theta_1 \right) \nn \\
& = \mathbb{E} \left[ \left. e^{\jmath \omega \gamma_m} \right|
v_1,\theta_1 , \gamma_m > \nu \right] \nn \\
& = \int_{\nu}^{\infty} e^{\jmath \omega v_m}
f_{\gamma_m | v_1,\theta_1} \left( v_m \right)
\text{d} v_m \nn \\
& = \sum_{n=0}^{\infty} d_n \left( v_1,\theta_1 \right)
\int_{\nu}^{\infty} v_m^n
e^{- \left( \zeta - \jmath \omega \right) v_m} \text{d} v_m \nn \\
& = \sum_{n=0}^{\infty} d_n \left( v_1,\theta_1 \right)
\frac{\Gamma \big(n+1, \left( \zeta - \jmath \omega \right) \nu \big)}
{\left( \zeta - \jmath \omega \right)^{n+1}} .
\label{eq64}
\end{align}
Utilizing the series-form expansion of the upper incomplete gamma function allows to rewrite \eqref{eq64} as
\begin{align}
& \! \! \! \Psi_{\gamma_m | v_1,\theta_1}
\left(\jmath \omega ; \nu, v_1,\theta_1 \right) \nn \\
& = e^{-\zeta \nu} e^{\jmath \omega \nu}
\sum_{n=0}^{\infty} d_n \left( v_1,\theta_1 \right) n!
\sum_{\ell=0}^{n} \frac{\nu^{\ell}}
{\ell! \left( \zeta - \jmath \omega \right)^{\left(n+1-\ell \right)}} .
\label{eq65}
\end{align}
{\color{black}{Since the ports contributing to ${\mathcal{H}}_{\textrm{a}}$ (respectively ${\mathcal{H}}_{\textrm{b}}$) are conditionally independent given $(v_1,\theta_1)$, the CF of their sum is the $K$-th (respectively $(K-1)$-th) power of \eqref{eq65}. Hence, for $J \in \{K,K-1\}$, we have
\begin{align}
& \! \! \! \! \! \! \! \!
\Psi_{J} \left( \jmath \omega ; \nu, v_1,\theta_1 \right)
= \Psi_{\gamma_m | v_1,\theta_1}^{J}
\left( \jmath \omega ; \nu, v_1,\theta_1 \right) \nn \\
& = e^{-\zeta J \nu} e^{\jmath \omega J \nu}
\sum_{n_1=0}^{\infty} \ldots \sum_{n_J=0}^{\infty}
\left( \prod_{m=1}^{J} d_{n_m} \left( v_1,\theta_1 \right) n_m! \right) \nn \\
& \quad \times \sum_{\ell_1=0}^{n_1} \ldots \sum_{\ell_J=0}^{n_J}
\left( \prod_{m=1}^{J} \frac{\nu^{\ell_m}}{\ell_m!} \right)
\left( \zeta - \jmath \omega \right)^{-\eta_J \left( \mathbf{n}, \boldsymbol{\ell} \right)} ,
\label{eq66}
\end{align}
with $\eta_J (\mathbf{n},\boldsymbol{\ell}) = J + \sum_{m=1}^{J} n_m - \sum_{m=1}^{J} \ell_m$ (cf.~\eqref{eq38}). Following similar steps as in \eqref{eq40}, we obtain the expression of the corresponding PDF as
\begin{align}
& \! \! \! f_{J} \left(u; \nu, v_1,\theta_1 \right)
= e^{-\zeta J \nu} \!
\sum_{n_1=0}^{\infty} \ldots \sum_{n_J=0}^{\infty}
\left( \prod_{m=1}^{J} \! d_{n_m} (v_1,\theta_1)  n_m! \! \right) \nn \\
& \! \! \! \times \sum_{\ell_1=0}^{n_1} \ldots \sum_{\ell_J=0}^{n_J}
\left( \prod_{m=1}^{J} \frac{\nu^{\ell_m}}{\ell_m!} \right)
\frac{\left(u - J \nu \right)^{\eta_J \left( \mathbf{n}, \boldsymbol{\ell} \right)-1}}
{\big( \eta_J \left( \mathbf{n}, \boldsymbol{\ell} \right) - 1 \big)!}
e^{-\zeta \left( u-J \nu \right)} , \nn \\
& \hspace{6.5cm} u > J \nu ,
\label{eq67}
\end{align}
using which the expressions of ${\mathcal{H}}_{\textrm{a}}$ and ${\mathcal{H}}_{\textrm{b}}$ can be written, for $J=K$ and $J=K-1$, respectively, as
\begin{align}
& \! \! \! {\mathcal{H}}_{a/b} \left( x, \nu, v_1,\theta_1 \right)
= \int_{J\nu}^{x}
f_{J} \left(u; \nu, v_1,\theta_1 \right) \, \text{d} u \nn \\
& = e^{-\zeta J \nu}
\sum_{n_1=0}^{\infty} \ldots \sum_{n_J=0}^{\infty}
\left( \prod_{m=1}^{J} d_{n_m} (v_1,\theta_1) \, n_m! \right) \nn \\
& \times
\sum_{\ell_1=0}^{n_1} \ldots \sum_{\ell_J=0}^{n_J}
\left( \prod_{m=1}^{J} \frac{\nu^{\ell_m}}{\ell_m!} \right)
\frac{\gamma \big( \eta_J \left( \mathbf{n},\boldsymbol{\ell} \right),
\zeta ( x - J \nu ) \big)}
{\big( \eta_J \left( \mathbf{n},\boldsymbol{\ell} \right)-1 \big)!\,
\zeta^{\eta_J \left( \mathbf{n},\boldsymbol{\ell} \right)}} ,
\label{eq75}
\end{align}
with ${\mathcal{H}}_{a/b} (x,\nu,v_1,\theta_1) = 0$ for $x < J\nu$ and where the target $\gamma_{\text{th}}$ for ${\mathcal{H}}_{\textrm{a}}$ is replaced by $\gamma_{\text{th}}-v_1$ for ${\mathcal{H}}_{\textrm{b}}$ (cf. \eqref{eqD3}). Substituting \eqref{eq75} (with $J=K$, $U=U'$, and $x=\gamma_{\text{th}}$ for ${\mathcal{H}}_{\textrm{a}}$, and $J=K-1$, $U=U''$, and $x=\gamma_{\text{th}}-v_1$ for ${\mathcal{H}}_{\textrm{b}}$) into \eqref{eqD2}--\eqref{eqD4} gives the OP conditioned on $(v_1,\theta_1)$ jointly. Finally, substituting \eqref{eq59} into \eqref{eq56} and averaging over $\theta_1$ and $\gamma_1$ results in the expression of the OP for the physical reference port model, given by
\begin{align}
P_{\text{out}}^{\text{p.r.p.}} & = \int_{0}^{\infty} \! \! \! \int_0^{2\pi}
\bigg( {{M-1} \choose {K}} (M-K-1)
\! \! \! \int\limits_{v_1}^{\gamma_{\text{th}}/K} \! \! \!
f_{\gamma_m | v_1,\theta_1}(\nu) \nn\\
& \quad \times F^{M-K-2}(\nu,v_1,\theta_1)
{\mathcal{H}}_{\textrm{a}} \left(\gamma_{\text{th}},\nu,v_1,\theta_1 \right) \text{d} \nu \nn \\
& \quad + {{M-1} \choose {K-1}} (M-K) \int_{0}^{v_1}
{\mathcal{H}}_{\textrm{b}} \left(\gamma_{\text{th}}-v_1,\nu,v_1,\theta_1 \right) \nn \\
& \quad \times
F^{M-K-1}(\nu,v_1,\theta_1) f_{\gamma_m | v_1,\theta_1}(\nu) \text{d} \nu \nn \\
& \quad + {{M-1} \choose {K}} F^{M-1-K}(v_1,v_1,\theta_1) \nn \\
& \quad \times {\mathcal{H}}_{\textrm{a}} \left(\gamma_{\text{th}},v_1,v_1,\theta_1 \right) \bigg)
f(\theta_1 \! \! \mid \! \! \gamma_1=v_1) \, \text{d} \theta_1 \, f_{\gamma_1}(v_1) \, \text{d}v_1 .
\label{eq76}
\end{align}

{\em Case of $K=M$:} When the FA receiver selects all ports, the event of outage occurs only from the term in \eqref{eqD1} with $U''=M-K-1=-1$, implying that \eqref{eq76} reduces to
\begin{align}
P_{\text{out} | K=M}^{\text{p.r.p.}}
\left( \gamma_{\text{th}} \right)
& = \int_{0}^{\gamma_{\text{th}}} \int_0^{2\pi}
{\mathcal{H}}_{\textrm{b}} \left(\gamma_{\text{th}}-v_1,0,v_1,\theta_1 \right) \nn \\
& \quad \times f(\theta_1 \mid v_1) \, \text{d}\theta_1 \,
f_{\gamma_1} \left( v_1 \right) \text{d} v_1 .
\label{eq77}
\end{align}
Setting $\nu=0$ and $J=M-1$ in \eqref{eq75}, the expression of ${\mathcal{H}}_{\textrm{b}} \left(\gamma_{\text{th}}-v_1,0,v_1,\theta_1 \right)$ simplifies to
\begin{align}
& \! \! \! \! \! {\mathcal{H}}_{\textrm{b}} \left(\gamma_{\text{th}}-v_1,0,v_1,\theta_1 \right) \nn \\
& \quad = \sum_{n_1=0}^{\infty} \ldots \sum_{n_{M-1}=0}^{\infty}
\left( \prod_{m=1}^{M-1} d_{n_m} \left( v_1,\theta_1 \right) \, n_m! \right) \nn \\
& \quad \times
\frac{\gamma \big( M-1 + \sum_{m=1}^{M-1} n_m ,
\zeta ( \gamma_{\text{th}}-v_1 ) \big)}
{\big( M-1 + \sum_{m=1}^{M-1} n_m - 1 \big)! \,
\zeta^{M-1 + \sum_{m=1}^{M-1} n_m}} .
\label{eq83}
\end{align}
Utilizing the relation $\sum_{n_1+\ldots+n_{M-1}=n} \frac{1}{\prod_{m=1}^M n_m!} = \frac{(M-1)^n}{n!}$ and from \eqref{eq22}, \eqref{eq83} simplifies to
\begin{align}
& \! \! \! \!
{\mathcal{H}}_{\textrm{b}} \left(\gamma_{\text{th}}-v_1,0,v_1,\theta_1 \right)
= \sum_{n=0}^{\infty} \frac{\big(\zeta (M-1) \Delta (v_1,\theta_1) \big)^{n}}{n!} \nn \\
& \quad \times e^{- \zeta (M-1) \Delta (v_1,\theta_1)}
\frac{\gamma \big( M-1+n, \zeta ( \gamma_{\text{th}}-v_1 ) \big)}
{\left( M-2+n \right)!} .
\label{eq83new}
\end{align}
Furthermore, from \eqref{eq17} and \eqref{eqC1}, we have
\beq
f_{\gamma_1}(v_1)\, f(\theta_1 \mid v_1) = \frac{\zeta (1 - \rho^2)}{2\pi}\, e^{-\kappa - \zeta (1 - \rho^2) v_1} e^{\xi(v_1)\cos\theta_1} .
\label{eqjoint}
\eeq
Using the definition of $\Delta(v_1,\theta_1)$ in \eqref{eqC2}, we apply algebraic simplifications using $\big(A(v_1) + B(v_1) \cos \theta_1\big)^n = \sum_{k=0}^n {{n}\choose{k}} A^{n-k} (v_1) B^k(v_1) \cos^k \theta_1$, the power-reduction identity $\cos^k \theta = 2^{-k} \sum_{i=0}^{k} {{k}\choose{i}} e^{\jmath(k-2i) \theta}$, the Jacobi-Anger expansion $e^{x \cos \theta} = \sum_{k = -\infty}^{\infty} I_k(x) e^{\jmath k \theta}$, and the orthogonality of $\{e^{\jmath k\theta}\}$ over one period, i.e.,
\begin{align}
\frac{1}{2\pi} \int_0^{2\pi} \cos^k \theta \, e^{x \cos \theta}\, \text{d} \theta
= 2^{-k} \sum_{i=0}^{k} {{k}\choose{i}} I_{k-2i}(x).
\end{align}
Following these steps, the final OP expression with $K=M$ for the FAS with the physical reference port model is obtained as
\begin{align}
P_{\text{out} | K=M}^{\text{p.r.p.}}
& = \int_0^{\gamma_{\text{th}}} \zeta (1-\rho^2) e^{-\kappa - \zeta (1-\rho^2) v_1 - \zeta(M-1) A(v_1)} \nn \\
& \quad \times \sum_{n=0}^{\infty}
\frac{\big(\zeta(M-1)\big)^n \gamma \big( M-1+n, \beta ( \gamma_{\text{th}}-v_1 ) \big)}{n! \left( M-2+n \right)!} \nn \\
& \quad \times \sum_{k=0}^{n} {{n}\choose{k}} A^{n-k}(v_1) \frac{B^k(v_1)}{2^k}
\sum_{i=0}^{k} {{k}\choose{i}} \nn \\
& \quad \times I_{k-2i} \left( C_{\zeta(M-1)} (v_1) \right) \text{d} v_1 ,
\label{eqKMphys}
\end{align}
with $C_p(v_1) = \xi (v_1) - p \zeta B (v_1)$.
}}
\section{Lower Bound and Asymptotic Analysis} \label{sec:lbaa}
The exact expressions of the OP in \eqref{eq46} and \eqref{eq76}, obtained for the virtual and physical reference port models, respectively, involve nested integrals and infinite series, and thus become mathematically intractable. Motivated by this, we derive more tractable lower bounds on the OP for the FAS with the virtual and physical reference port models in Sections~\ref{sec:OP_bound_virtual} and~\ref{sec:OP_bound_physical}, respectively.
\subsection{OP's Lower Bound with the Virtual Reference Port Model} \label{sec:OP_bound_virtual}
Considering the virtual reference port model, the lower bound on the OP in \eqref{eq46} is derived by lower-bounding the expressions of ${\mathcal{G}} \left( \nu, v_0 \right)$ and ${\mathcal{H}} \left(\gamma_{\text{th}}, \nu, v_0 \right)$ in \eqref{eq31} and \eqref{eq45}, respectively. First, using $I_0(x)\ge 1$, $f_{\gamma_m | Y_0 = y_0} \left(v_m \right)$ in \eqref{eq11} can be lower-bounded as
\beq
f_{\gamma_m | Y_0 = y_0}^{\text{LB}} \left(v_m \right)
= \zeta \, e^{-\zeta \left( v_m + y_0 \right)}, \
v_m \geq 0 ,
\label{eq86}
\eeq
which yields the lower bound on ${\mathcal{G}} \left( \nu, v_0 \right)$ given by
\begin{align}
{\mathcal{G}}^{\text{LB}} \left( \nu, y_0 \right)
& = \prod_{m \in {\bar{\mathcal{S}}}} \int_{0}^{\nu}
f_{\gamma_m | Y_0=y_0}^{\text{LB}} \left(v_m\right)
\text{d} v_m \nn \\
& = e^{-U \zeta y_0}
\left(1-e^{-\zeta \nu}\right)^U .
\label{eq87}
\end{align} 
Similarly, using \eqref{eq86}, the CF expression in \eqref{eq36} can be lower-bounded as
\begin{align}
\Psi_{\sum\limits_{m \in {\mathcal{S}}} \gamma_m | Y_0}^{\text{LB}}
\left( \jmath \omega ; \nu, y_0 \right)
& = \left( \int_{\nu}^{\infty} \! \! \! e^{\jmath \omega v_m}
f_{\gamma_m | Y_0=y_0}^{\text{LB}} \left(v_m\right)
\text{d} v_m \right)^K \nn \\
& = \left( \frac{\zeta}{\zeta - \jmath \omega} \right)^K \! \!
e^{-K \zeta y_0}
e^{- \left( \zeta - \jmath \omega \right) K \nu} ,
\label{eq88}
\end{align}
which results in the lower bound on the PDF in \eqref{eq41} given by
\begin{align}
f_{\sum\limits_{m \in \mathcal{S}} \gamma_m ,
\gamma_m > \nu | Y_0}^{\text{LB}} \left( u \right)
& = e^{-K \zeta y_0}
\frac{e^{-\zeta u} \zeta^{K}}{\left( K-1 \right)!}
\left(u - K \nu \right)^{K-1} , \nn \\
& \hspace{3.2cm} u > K \nu .
\label{eq89}
\end{align}
Accordingly, a lower bound on ${\mathcal{H}} \left(\gamma_{\text{th}}, \nu, y_0 \right)$ can be derived as
\begin{align}
{\mathcal{H}}^{\text{LB}} \left(\gamma_{\text{th}}, \nu, y_0 \right)
& = e^{-K\zeta y_0} e^{-\zeta K \nu}
\frac{\gamma \big(K, \zeta ( \gamma_{\text{th}} - K \nu ) \big)}{\left( K-1 \right)!}, \nn \\
& \hspace{4cm} \gamma_{\text{th}} \ge K \nu .
\label{eq90}
\end{align} 

{\color{black}{
Substituting \eqref{eq86}--\eqref{eq87} and \eqref{eq90} into \eqref{eq29}, the lower bound on the OP for the virtual reference port model is
\begin{align}
P_{\text{out}}^{\text{v.r.p., LB}}
& = {{M}\choose{K}} (M-K) \mathbb{E}_{Y_0} \left[ e^{-\zeta M y_0} \right]  
\int\limits_0^{\gamma_{\text{th}}/K} \! \! \zeta e^{-\zeta(K+1)\nu} \nn \\
& \quad \times \left(1-e^{-\zeta\nu}\right)^U
\frac{\gamma\big(K,\zeta(\gamma_{\text{th}}-K\nu)\big)}{(K-1)!} \text{d}\nu .
\label{eqE1}
\end{align}
Moreover, using \eqref{eq8}, we have
\begin{align}
\mathbb{E}_{Y_0} \left[ e^{-\zeta M y_0} \right] &\!= \!\int_0^{\infty} e^{-\zeta M y_0} f_{Y_0} (y_0) \text{d} y_0 \nn \\
&= \! -\zeta_0 e^{-\zeta_0 \delta}
\! \! \int_0^{\infty} \! \! \! \! e^{-(\zeta_0 + M \zeta)y_0} I_0 \big( 2\zeta_0 \sqrt{\delta y_0}\big) \text{d}y_0 \nn \\
&= \! \frac{1-\rho^2}{1+(M-1)\rho^2} e^{-\frac{M \kappa}{1+(M-1)\rho^2}} .
\label{eq:exp_y0_lbvrp}
\end{align}
Now, substituting \eqref{eq:exp_y0_lbvrp} into \eqref{eqE1} and employing the series-form expansion of the lower incomplete gamma function, i.e.,
\begin{align}
\frac{\gamma \left(K,x \right)}{\left(K-1 \right)!}
= 1-e^{-x}\sum_{k=0}^{K-1}\frac{x^k}{k!},
\end{align}
we can express \eqref{eqE1} as
\begin{align}
{P}_{\text{out}}^{\text{v.r.p., LB}}
& = \frac{(1-\rho^2)(M-K)}{1+(M-1)\rho^2} e^{-\frac{M \kappa}{1+(M-1)\rho^2}} {{M} \choose {K}} \nn \\
& \quad \times \int_0^{\gamma_{\text{th}}/K} \zeta e^{-\zeta(K+1)\nu}
\left( 1-e^{-\zeta\nu} \right)^U \nn \\
& \quad \times \left( 1 - e^{-\zeta(\gamma_{\text{th}}-K\nu)} \sum_{k=0}^{K-1}
\frac{\big(\zeta(\gamma_{\text{th}}-K\nu)\big)^k}{k!} \right) \text{d}\nu .
\label{eq95}
\end{align}
For the case of $K=M$, the lower bound further simplifies to
\beq
P_{\text{out} | K=M}^{\text{v.r.p., LB}}
= \frac{(1-\rho^2)\gamma \left(M, \zeta \gamma_{\text{th}} \right)}
{\big(1+(M-1)\rho^2\big)(M-1)!} e^{-\frac{M \kappa}{1+(M-1)\rho^2}} .
\label{eqE3}
\eeq
}}
\subsection{OP's Lower Bound with the Physical Reference Port Model} \label{sec:OP_bound_physical}
{\color{black}{Considering the physical reference port model, the lower bound on the OP in \eqref{eq76} is derived by lower-bounding the PDF $f_{\gamma_m | v_1, \theta_1} \left(v_m \right)$ in \eqref{eq19} and considering the three cases of port selection for the physical reference port model described in Section~\ref{sec:OP_physical}. To this end, we introduce the set of parameters corresponding to the three cases as
\beq
(J, U, x, {\mathcal{C}} ) =
\begin{cases}
\big(K,\; U',\; \gamma_{\text{th}},\; {M-1 \choose K}(M-K-1)\big), & \\
& \hspace{-1.6cm} \text{case (a)}, \\
\big(K-1,\; U'',\; \gamma_{\text{th}}-v_1,\; {M-1 \choose K-1}(M-K)\big), & \\
& \hspace{-1.6cm} \text{case (b)}, \\
\big(K,\; U'',\; \gamma_{\text{th}},\; {M-1 \choose K}\big), & \hspace{-1.6cm} \text{case (c)},
\end{cases}
\label{eqDefs}
\eeq
where $J$ is the number of selected non-reference ports, $U$ is the number of non-reference ports below the boundary port, $x$ denotes the threshold for their sum, and $\mathcal{C}$ is a combinatorial factor denoting the number of ways to assign the roles of boundary, selected, and unselected ports to the $M-1$ non-reference ports in the respective case. In addition, we introduce the integration range of $\nu$ for the three cases as
\beq
\mathcal{R} =
\begin{cases}
\big[v_1,\ \gamma_{\text{th}}/K\big], &
\hspace{-1.6cm} \text{case (a), for } 0\le v_1\le \gamma_{\text{th}}/K , \\
\big[0,\ \tilde v_1\big],\ \tilde v_1=\min\!\big\{v_1,\tfrac{\gamma_{\text{th}}-v_1}{K-1}\big\}, & \\
& \hspace{-1.6cm} \text{case (b), for } 0\le v_1\le \gamma_{\text{th}} ,\\
\{v_1\}, & \hspace{-1.6cm} \text{case (c), for } 0\le v_1\le \gamma_{\text{th}}/K.
\end{cases}
\label{eqRanges}
\eeq

To compute the lower bound of $f_{\gamma_m | v_1, \theta_1} \left(v_m \right)$ we use $I_0(\cdot) \ge 1$ to get
\beq
f_{\gamma_m | \gamma_1=v_1}^{\text{LB}} \left(v_m \right)
= \zeta e^{-\tau\kappa} e^{-\zeta \left(v_m + \rho^{2} v_1 \right)}
\frac{I_0 \big(C_1(v_1) \big)}{I_0 \big(\xi(v_1) \big)} ,
\ v_m \ge 0 .
\label{eq99}
\eeq
The corresponding lower bound on $F^U(\nu,v_1,\theta_1)$ can be expressed as
\begin{align}
\! \! \!
F^{U}_{\rm LB} (\nu,v_1,\theta_1)
= e^{-U \zeta \Delta (v_1,\theta_1)}
\sum_{u=0}^{U} \binom{\! \! U \! \!}{\! \! u \! \!}
\left( -1 \right)^u e^{- \zeta u \nu} .
\label{eq100}
\end{align}
Similarly, using \eqref{eq99}, the CF expression in \eqref{eq66} can be lower-bounded (see \eqref{eq58}--\eqref{eqD1}) as
\beq
\Psi_{J}^{\text{LB}}
\left( \jmath \omega ; \nu, v_1,\theta_1 \right)
= \zeta^{J} e^{- J \zeta \Delta (v_1,\theta_1)}
\frac{e^{- J \left( \zeta - \jmath \omega \right) \nu}}
{\left( \zeta - \jmath \omega \right)^{J}} ,
\label{eq101}
\eeq
which results in the lower bound on the PDF of the sum of $J$ such ports, restricted to exceed $\nu$, given by
\begin{align}
f^{\text{LB}}_{J} \! \! \left(u; \nu,v_1,\theta_1 \right)
& = \zeta^J e^{- J \zeta \Delta (v_1,\theta_1)}
e^{-\zeta u} \frac{ \left(u- J \nu \right)^{J-1}}{\left( J-1 \right)!} , \nn \\
& \hspace{4cm} u > J \nu .
\label{eq104}
\end{align}
Accordingly, a lower bound on ${\mathcal{H}}_{\textrm{a}/\textrm{b}} \left(x, \nu, v_1,\theta_1 \right)$ (with $J$ and $x$ given in \eqref{eqDefs}) can be derived as
\begin{align}
{\mathcal{H}}_{{\textrm{a}/\textrm{b}}}^{\text{LB}} \left(x, \nu, v_1,\theta_1 \right)
& = e^{- J \zeta \Delta (v_1,\theta_1)}
e^{-\zeta J \nu}
\frac{\gamma \big( J, \zeta(x - J \nu)\big)}
{\left(J -1 \right)!} , \nn \\
& \hspace{4cm} x \geq J \nu .
\label{eq103}
\end{align}
Furthermore, a lower bound on the term in \eqref{eqD4} follows the same substitution, with $F^{\text{LB}}$ in \eqref{eq100} and ${\mathcal{H}}_{\textrm{a}}^{\text{LB}}$ in \eqref{eq103} evaluated at $\nu=v_1$, and is computed as
\begin{align}
P_{\textrm{c}}^{\text{LB}} \left(\gamma_{\text{th}}, v_1,\theta_1 \right)
& = {{M-1} \choose {K}} F_{\rm LB}^{M-1-K}(v_1,v_1,\theta_1) \nn \\
& \quad \times 
{\mathcal{H}}_{\textrm{a}}^{\text{LB}} \left(\gamma_{\text{th}}, v_1, v_1,\theta_1 \right) .
\label{eq103c}
\end{align}

Substituting \eqref{eq100} and \eqref{eq103} into \eqref{eqD2}--\eqref{eqD4},
we have
\begin{align}
P_{\textrm{a}}^{\text{LB}} + P_{\textrm{b}}^{\text{LB}} + P_{\textrm{c}}^{\text{LB}} 
& = e^{-(M-1) \zeta \Delta (v_1,\theta_1)} \nn \\
& \quad \times \left( \bar{P}_{a,v_1}^{\text{LB}} + \bar{P}_{b,v_1}^{\text{LB}} + \bar{P}_{c,v_1}^{\text{LB}} \right) ,
\label{eq106}
\end{align}
where $\bar{P}_{a,v_1}^{\text{LB}}$, $\bar{P}_{b,v_1}^{\text{LB}}$, and $\bar{P}_{c,v_1}^{\text{LB}}$ are defined as the $\theta_1$-independent factors of \eqref{eqD2}--\eqref{eqD4} that remain after extracting the term $e^{-(M-1) \zeta \Delta (v_1,\theta_1)}$ in \eqref{eq106}.
With the parameters given in \eqref{eqDefs}, these factors are expressed as
\begin{align}
\bar P_{i} & = {\mathcal{C}} \int_{\mathcal{R}} \zeta e^{-\zeta (J+1) \nu} \left(1-e^{-\zeta \nu}\right)^{U}
\frac{\gamma \big(J,\zeta(x-J\nu)\big)}{(J-1)!} \, \text{d}\nu , \nn \\
& \hspace{6cm} i \in \{\textrm{a},\textrm{b}\} ,
\label{eqPbarab} \\
\bar P_{\textrm{c}} & = {\mathcal{C}} \, e^{-\zeta K v_1} \left(1-e^{-\zeta v_1}\right)^{U}
\frac{\gamma \big(K,\zeta(\gamma_{\text{th}}-K v_1)\big)}{(K-1)!} ,
\label{eqPbarc}
\end{align}
where each factor is equal to $0$ outside the integration range of $v_1$ in \eqref{eqRanges}. Additionally, we have $\bar P_{\textrm{a}}=0$ for $K=M-1$, $\bar P_{\textrm{c}}=0$ for $K=M$, and $\bar P_{\textrm{b}}=\left(1-e^{-\zeta v_1}\right)^{M-1}$ for $0\le v_1\le\gamma_{\text{th}}$ and $K=1$ (with ${\mathcal{H}}_{\textrm{b}}=1$).


Using the definition of $\Delta(v_1, \theta_1)$ in \eqref{eqC2} along with \eqref{eqC1}, we have
\beq
\mathbb{E}_{\theta_1 \mid v_1} \! \! \big[ e^{-(M-1) \zeta \Delta (v_1,\theta_1)} \big]
\! = \! e^{-(M-1) \zeta A (v_1)} \frac{I_0 \big( C_{M-1}(v_1) \big)}{I_0 \big( \xi(v_1) \big)}.
\label{eqCM}
\eeq
Averaging \eqref{eq106} first over $\theta_1$ using \eqref{eqCM} and then over $\gamma_1$ using $f_{\gamma_1}(v_1)=\zeta_1e^{-\kappa-\zeta_1v_1}I_0\big(\xi(v_1)\big)$ (cf. \eqref{eq17}), the lower bound on the OP for the physical reference port model is
\begin{align}
P_{\text{out}}^{\text{p.r.p., LB}}
& = \int_{0}^{\gamma_{\text{th}}} \! \! \! \! \zeta_1 e^{-\kappa-\zeta_1 v_1}
I_0 \big(C_{M-1}(v_1)\big)
e^{-(M-1)(\tau \kappa + \beta \rho^2 v_1)} \nn \\
& \qquad \quad \times \left( \bar{P}_{a,v_1}^{\text{LB}} + \bar{P}_{b,v_1}^{\text{LB}} + \bar{P}_{c,v_1}^{\text{LB}} \right)
\text{d} v_1 ,
\label{eq107}
\end{align}
with $\tau = (1-\rho)/(1+\rho)$. 

For the case of $K=M$, cases~(a) and (c) described in Section~\ref{sec:OP_physical} do not contribute to the OP. Thus, the lower bound in \eqref{eq107} is simplified by considering the bound only for case~(b), which is obtained by setting $J=M-1$, $\nu=0$, and $x=\gamma_{\text{th}}-v_1$ in \eqref{eq103} and averaging over $\theta_1$ (similar to \eqref{eqCM}). Furthermore, averaging over $\gamma_1$ using \eqref{eq17}, we obtain
\begin{align}
P_{\text{out} | K=M}^{\text{p.r.p., LB}}
& = \! \int_{0}^{\gamma_{\text{th}}} \! \zeta_1 e^{-\kappa-\zeta_1 v_1}
I_0 \left(C_{M-1}(v_1)\right) \nn \\
& \quad \times e^{-(M-1)(\tau \kappa + \zeta \rho^2 v_1)}
\frac{\gamma \big(M \! - \! 1, \zeta(\gamma_{\text{th}}-v_1) \big)}{(M-2)!}
\text{d} v_1 .
\label{eq107e}
\end{align}
\subsection{Asymptotic OP Analysis at High Average SNR}
Considering the virtual reference port model, a high average SNR $\bar{\gamma} \gg 1$ implies $\zeta = \left(\bar{\gamma}(1-\rho^{\color{black}2})\sigma_h^{2} \right)^{-1} \ll 1$. For $v\le\gamma_{\text{th}}$ and fixed $y_0$, \eqref{eq12} gives $f_{\gamma_m\mid Y_0=y_0}(v) \to d_0(y_0) = \zeta e^{-\zeta y_0}$, as $I_0(2\zeta\sqrt{v y_0})\to 1$ and $e^{-\zeta v}\to 1$ for $\zeta v\ll1$. Since the $M$ ports are conditionally i.i.d. given $Y_0$, the conditional probability in \eqref{eq26} is obtained as
\begin{align}
\Pr \! \left( \left. \sum\limits_{m \in \mathcal{S}} \gamma_m \le \gamma_{\text{th}} \right| Y_0 \! = \! y_0 , \bar{\gamma} \gg 1 \right)
& \to \frac{d_0^M(y_0) \left( \zeta \gamma_{\text{th}} \right)^M}{K! \, K^{M-K}} \nn \\
& = \frac{e^{-M\zeta y_0}\, \left( \zeta \gamma_{\text{th}} \right)^M}{K! \, K^{M-K}} .
\label{eq97vol}
\end{align}
Utilizing $\mathbb{E}_{Y_0}[e^{-M\zeta Y_0}]$ from \eqref{eq:exp_y0_lbvrp}, we obtain the asymptotic expression of the OP as
\beq
P_{\text{out} | \bar{\gamma} \gg 1 }^{\text{v.r.p.}}
\approx C^{\text{v.r.p.}} \left( \zeta \gamma_{\text{th}} \right)^{M} ,
\label{eq97}
\eeq
with 
\beq
C^{\text{v.r.p.}}
= \frac{1-\rho^2}{K! \, K^{M-K} \big((M-1) \rho^2 + 1\big)}
\, e^{- \frac{M\kappa}{M\rho^2-\rho^2+1}} .
\label{eq98}
\eeq

Similarly, considering the physical reference port model, $\bar{\gamma} \gg 1$ and $\zeta \ll 1$ imply that an outage requires small SNR at all $M$ ports simultaneously, including the reference port (port~$1$), which is also a candidate for selection. Expressing $y_1 = \zeta v_1$ and assuming $y_1 \ll 1$ (which implies $v_1 \ll 1$ for $\zeta \ll 1$), the marginal density of $y_1$ in \eqref{eq17} tends to the constant $f_{Y_1}(0) = (1-\rho^2) e^{-\kappa}$, while the conditional density \eqref{eq21} of each of the remaining $M-1$ ports, given $y_1 \to 0$, tends to the constant $d_0(0) = e^{-\kappa \tau}$ (due to $B (v_1) \to 0$ and $A (v_1) \to \kappa(1-\rho)^2/(1-\rho^2) = \kappa \tau$ for $v_1 \to 0$, resulting in $\xi(v_1) \to 0$). This leads to the asymptotic expression of the OP as
\beq
P_{\text{out} | \bar{\gamma} \gg 1 }^{\text{p.r.p.}}
\approx C^{\text{p.r.p.}} \left( \zeta \gamma_{\text{th}} \right)^{M} ,
\label{eq109}
\eeq
with
\beq
C^{\text{p.r.p.}} = \frac{(1-\rho^2) \, e^{-\kappa} \, e^{-(M-1)\tau \kappa}}{K! \, K^{M-K}}.
\label{eq110}
\eeq

We can make the following observations from \eqref{eq97}--\eqref{eq110}:
\begin{itemize}
\item From \eqref{eq97} and \eqref{eq109}, using $\zeta = (\kappa+1)/\left(\bar\gamma(1-\rho^2) \right)$, we have $P_{\text{out} | \bar{\gamma} \gg 1 }^{\text{v.r.p.}}, P_{\text{out} | \bar{\gamma} \gg 1 }^{\text{p.r.p.}} \propto (\bar{\gamma})^{-M}$ and, thus, the diversity order of the MRC-based FAS employing either of the reference port models is $M$. This holds even though only $K$ of the $M$ ports are combined, as an outage requires all $M$ ports to be weak, including the $M-K$ ports that are not combined.
\item The derived diversity order is valid for $\rho<1$ strictly, for which the port correlation matrix has full rank. In fact, for $\rho \to 1$, all $M$ ports share a single degree of freedom, and the diversity order correspondingly degrades to $1$ regardless of $M$ and $K$, consistent with \eqref{eq97} and \eqref{eq109}. The diversity order is still valid for small $W$, which leads to the approximation $1-\rho^2 \approx \pi^2W^2/6$, thus affecting only the coding gain.
\item For the case of independent ports, i.e., for $\rho=0$, the expressions in \eqref{eq97} and \eqref{eq109} reduce to $\big((\kappa+1)e^{-\kappa}\gamma_{\text{th}}/\bar{\gamma}\big)^M/(K!\,K^{M-K})$. For $K=1$, this is the $M$-th power of $(\kappa+1)e^{-\kappa}\gamma_{\text{th}}/\bar{\gamma}$, corresponding to the high-SNR approximation of the CDF of the SNR of a single Rician port evaluated at $\gamma_{\text{th}}$, as expected for selection among $M$ i.i.d. Rician ports.
\item For the case of Rayleigh fading, i.e., $\kappa = 0$, the exponential factors equal one, leading to $P_{\text{out}}\propto(\gamma_{\text{th}}/\bar{\gamma})^M$. For $K=M$, the factor $K!\,K^{M-K}$ simplifies to $M!$, recovering the classical MRC asymptote. Similarly, for $\kappa \to \infty$, the factors $e^{-M\kappa/D}$ and $e^{-\kappa[1+(M-1)\tau]}$ vanish exponentially, consistent with a channel approaching the deterministic LoS value $\bar{\gamma}\kappa/(\kappa+1)$ per port, for which there is no fading and thus no diversity.
\item Lastly, $K=1$ (maximal-gain combining) gives $K!\,K^{M-K}=1$, whereas $K=M$ (all ports combined) gives $M!$. Thus, the slope $M$ is the same for all $K \in [1, M]$, and $K$ only affects the coding gain.
\end{itemize}
}}
 \section{Numerical Results} \label{sec:nm}
This section presents numerical results to validate the analytical framework, study the impact of key system parameters on the performance of the considered MRC-based FAS, and compare the virtual and physical reference port models. \textcolor{black}{In the following, the cases of $K=1$ and $K=M$ correspond to maximal-gain combining and MRC with all ports combined, respectively.}

\textcolor{black}{Figs.~\ref{f1} and~\ref{f4} plot the OP versus the average SNR $\bar{\gamma}$ considering the virtual and physical reference port models, respectively, for different values of $M$ and $K$, $\gamma_{\text{th}}=5$~dB, $\kappa=0.2$, and $W=1$. The simulated OP in \eqref{eq25}, labeled `simul.', is obtained via Monte Carlo simulations by averaging over $10^{6}$ channels generated using \eqref{eq6} and \eqref{eq15} for the virtual and physical reference port models, respectively. This is compared with the analytical expressions in \eqref{eq46} and \eqref{eq76}, the lower bounds in \eqref{eq95} and \eqref{eq107}, and the asymptotic expressions at high average SNR in \eqref{eq97} and \eqref{eq109} for the virtual and physical reference port models, respectively.}
The system parameters are chosen to ensure reasonably low OP values for varying $M$ and $K$. For both figures, we observe that the analytical OP coincides with the simulated one, confirming the correctness of the analysis. As expected, the OP performance of the FAS also improves with $M$ and $K$. Moreover, the lower bound becomes increasingly tight as $\bar{\gamma}$ grows, and the asymptotic OP approaches the exact one at high SNR, confirming the diversity order of the FAS to be $M$ for both reference port models. \textcolor{black}{The lower bound turns out to be more tight for the virtual reference port model compared with the physical reference port model, while the tightness of the latter increases with $M$ and $K$.}

\textcolor{black}{In addition to these curves, Figs.~\ref{f1} and \ref{f4} also present the OP obtained via Monte Carlo simulations by averaging over $10^8$ channels generated using the fully correlated Jakes' model (denoted by `$P_{\text{out}}$ (Jakes')') and the distance-dependent physical reference port model (denoted by `$P_{\text{out}}^{\text{p.r.p., dist.}}$'), implying that the correlation coefficient $\rho$ in \eqref{eq15} is replaced by $\rho_m = J_0(2\pi(m-1)W/(M-1))$ for $ m \in \{ 2,\ldots,M\}$. It is observed that the curves obtained with Jakes' model exhibit a flatter slope at moderate SNR than those obtained with the two reference port models. This is because, for a fixed aperture $W$, the correlation matrix of Jakes' model has only about $2W+1$ significant eigenvalues \cite{9264694}, and additional ports provide limited independent fading. In contrast, the virtual and physical reference port models each retain $M-1$ eigenvalues equal to $1-\rho^2$, allowing additional ports to contribute more substantially to the performance gains. The diversity order $M$ is still attained by Jakes' model asymptotically, since its correlation matrix has full rank. However, this asymptotic regime requires $(\kappa+1)\gamma_{\text{th}}/\bar{\gamma}$ to fall below its smallest eigenvalue. For large $M$ and small $W$, the required SNR exceeds the plotted SNR range; at practical SNRs, the effective diversity is therefore governed by the number of eigenvalues appreciably larger than this ratio. These differences underscore the approximate nature of the reference port models. Nevertheless, the virtual reference port model more closely matches the distance-dependent physical reference port model compared with the physical reference port model, supporting its use as the more accurate of the two approximations.}

\begin{figure}[!t]
\centering
\begin{tikzpicture}

\begin{semilogyaxis}[
	width=3.4in, height=2.5in,
	xmin=0, xmax=15,
	ymin=1e-15, ymax=1e0,
	xlabel={Average SNR $\bar{\gamma}$ [dB]},
	ylabel={Outage probability},
    title={$\gamma_{\text{th}}=5$ dB, $\kappa=0.2$, $W=1$},
	xlabel near ticks,
	ylabel near ticks,
	label style={font=\footnotesize},
    xtick={0,1,...,15},
    ytick={1e-15,1e-12,1e-9,1e-6,1e-3,1e0},
    ticklabel style={font=\footnotesize},
	legend style={at={(0.01,0.01)}, anchor=south west, font=\scriptsize, inner sep=1pt, row sep=-2pt, fill opacity=0.6, draw opacity=1, text opacity=1},
	legend cell align=left,
	title style={font=\scriptsize, yshift=-2mm},
	grid=major
]

\addplot[thick, black, only marks, mark=asterisk, mark size=1.5pt]
table [x=SNR_dB, y=OP_simul, col sep=comma] {figures/files_txt/plot1new/op_snr_fas_vrp_5M_1K_5gamth_0.2kappa_1W.txt};
\addlegendentry{$P_{\text{out}}^{\text{v.r.p.}}$ (simul.)};

\addplot[thick, black, mark=o]
table [x=SNR_dB, y=OP_comp, col sep=comma] {figures/files_txt/plot1new/op_snr_fas_vrp_5M_1K_5gamth_0.2kappa_1W.txt};
\addlegendentry{$P_{\text{out}}^{\text{v.r.p.}}$ (Eq. \eqref{eq46})};

\addplot[thick, black, dashed]
table [x=SNR_dB, y=OP_LB, col sep=comma] {figures/files_txt/plot1new/op_snr_fas_vrp_5M_1K_5gamth_0.2kappa_1W.txt};
\addlegendentry{$P_{\text{out}}^{\text{v.r.p., LB}}$ (Eq. \eqref{eq95})};

\addplot[thick, black, densely dotted]
table [x=SNR_dB, y=OP_asymp, col sep=comma] {figures/files_txt/plot1new/op_snr_fas_vrp_5M_1K_5gamth_0.2kappa_1W.txt};
\addlegendentry{$P_{\text{out} | \bar{\gamma} \gg 1 }^{\text{v.r.p.}}$ (Eq. \eqref{eq97})};

\addplot[thick, black, mark=+]
table [x=SNR_dB, y=OP_simul, col sep=comma] {figures/files_txt/plot1new/op_snr_fas_jakes_5M_1K_5gamth_0.2kappa_1W.txt};
\addlegendentry{$P_{\text{out}}$ (Jakes')};

\addplot[thick, black, mark=x, mark size=1.5pt]
table [x=SNR_dB, y=OP_simul, col sep=comma] {figures/files_txt/plot1new/op_snr_fas_rhom_5M_1K_5gamth_0.2kappa_1W.txt};
\addlegendentry{$P_{\text{out}}^{\text{p.r.p., dist.}}$};
\addplot[thick, blue, only marks, mark=asterisk, mark size=1.5pt]
table [x=SNR_dB, y=OP_simul, col sep=comma] {figures/files_txt/plot1new/op_snr_fas_vrp_5M_4K_5gamth_0.2kappa_1W.txt};

\addplot[thick, blue, mark=o]
table [x=SNR_dB, y=OP_comp, col sep=comma] {figures/files_txt/plot1new/op_snr_fas_vrp_5M_4K_5gamth_0.2kappa_1W.txt};

\addplot[thick, blue, dashed]
table [x=SNR_dB, y=OP_LB, col sep=comma] {figures/files_txt/plot1new/op_snr_fas_vrp_5M_4K_5gamth_0.2kappa_1W.txt};

\addplot[thick, blue, densely dotted]
table [x=SNR_dB, y=OP_asymp, col sep=comma] {figures/files_txt/plot1new/op_snr_fas_vrp_5M_4K_5gamth_0.2kappa_1W.txt};

\addplot[thick, blue, mark=+]
table [x=SNR_dB, y=OP_simul, col sep=comma] {figures/files_txt/plot1new/op_snr_fas_jakes_5M_4K_5gamth_0.2kappa_1W.txt};

\addplot[thick, blue, mark=x, mark size=1.5pt]
table [x=SNR_dB, y=OP_simul, col sep=comma] {figures/files_txt/plot1new/op_snr_fas_rhom_5M_4K_5gamth_0.2kappa_1W.txt};
\addplot[thick, red, only marks, mark=asterisk, mark size=1.5pt]
table [x=SNR_dB, y=OP_simul, col sep=comma] {figures/files_txt/plot1new/op_snr_fas_vrp_10M_2K_5gamth_0.2kappa_1W.txt};

\addplot[thick, red, mark=o]
table [x=SNR_dB, y=OP_comp, col sep=comma] {figures/files_txt/plot1new/op_snr_fas_vrp_10M_2K_5gamth_0.2kappa_1W.txt};

\addplot[thick, red, dashed]
table [x=SNR_dB, y=OP_LB, col sep=comma] {figures/files_txt/plot1new/op_snr_fas_vrp_10M_2K_5gamth_0.2kappa_1W.txt};

\addplot[thick, red, densely dotted]
table [x=SNR_dB, y=OP_asymp, col sep=comma] {figures/files_txt/plot1new/op_snr_fas_vrp_10M_2K_5gamth_0.2kappa_1W.txt};
\addplot[thick, green!50!black, only marks, mark=asterisk, mark size=1.5pt] 
table [x=SNR_dB, y=OP_simul, col sep=comma] {figures/files_txt/plot1new/op_snr_fas_vrp_10M_10K_5gamth_0.2kappa_1W.txt};

\addplot[thick, green!50!black, mark=o]
table [x=SNR_dB, y=OP_comp, col sep=comma] {figures/files_txt/plot1new/op_snr_fas_vrp_10M_10K_5gamth_0.2kappa_1W.txt};

\addplot[thick, green!50!black, dashed]
table [x=SNR_dB, y=OP_LB, col sep=comma] {figures/files_txt/plot1new/op_snr_fas_vrp_10M_10K_5gamth_0.2kappa_1W.txt};

\addplot[thick, green!50!black, densely dotted]
table [x=SNR_dB, y=OP_asymp, col sep=comma] {figures/files_txt/plot1new/op_snr_fas_vrp_10M_10K_5gamth_0.2kappa_1W.txt};

\node[draw, thin, ellipse, minimum width=0.2cm, minimum height=0.5cm, inner sep=0pt] at (axis cs:12,6.5e-4) {};

\node[draw, thin, ellipse, minimum width=0.2cm, minimum height=0.55cm, inner sep=0pt] at (axis cs:10,1e-4) {};

\node[draw, thin, ellipse, minimum width=0.2cm, minimum height=0.4cm, inner sep=0pt] at (axis cs:11,2e-8) {};

\node[draw, thin, ellipse, minimum width=0.2cm, minimum height=0.3cm, inner sep=0pt] at (axis cs:13,5e-14) {};

\begin{scope}[>=latex]
\draw[->, thin] (12,4e-3) -- (12.25,3e-2);
\draw[->, thin] (9.8,1e-4) -- (4,6e-7);
\draw[->, thin] (10.8,2e-8) -- (8.7,2e-12);
\draw[->, thin] (12.8,5e-14) -- (10.9,1.5e-14);
\end{scope}
\node[black, font=\scriptsize] at (13,7e-2) {$M = 5, K=1$};

\node[black, font=\scriptsize] at (2,6e-7) {$M=5, K=4$};

\node[black, font=\scriptsize] at (8.7,6e-13) {$M=10, K=2$};

\node[black, font=\scriptsize] at (8.7,1.5e-14) {$M=10, K=10$};
\end{semilogyaxis}
\end{tikzpicture}
\caption{\textcolor{black}{OP versus average SNR $\bar{\gamma}$ considering the virtual reference port model, with $\gamma_{\text{th}}=5$~dB, $\kappa=0.2$, $W=1$, $M=\{5,10\}$, and $K=\{1,2,4,10\}$.}}
\label{f1}
\end{figure}

\begin{figure}[!t]
\centering
\begin{tikzpicture}

\begin{semilogyaxis}[
	width=3.4in, height=2.5in,
	xmin=0, xmax=15,
	ymin=1e-12, ymax=1e0,
	xlabel={Average SNR $\bar{\gamma}$ [dB]},
	ylabel={Outage probability},
    title={$\gamma_{\text{th}}=5$ dB, $\kappa=0.2$, $W=1$},
	xlabel near ticks,
	ylabel near ticks,
	label style={font=\footnotesize},
    xtick={0,1,...,15},
    ytick={1e-12,1e-10,1e-8,1e-6,1e-4,1e-2,1e0},
    ticklabel style={font=\footnotesize},
	legend style={at={(0.01,0.01)}, anchor=south west, font=\scriptsize, inner sep=1pt, row sep=-2pt, fill opacity=0.6, draw opacity=1, text opacity=1},
	legend cell align=left,
	title style={font=\scriptsize, yshift=-2mm},
	grid=major
]

\addplot[thick, black, only marks, mark=asterisk, mark size=1.5pt]
table [x=SNR_dB, y=OP_simul, col sep=comma] {figures/files_txt/plot2new/op_snr_fas_prp_5M_1K_5gamth_0.2kappa_1W.txt};
\addlegendentry{$P_{\text{out}}^{\text{p.r.p.}}$ (simul.)};

\addplot[thick, black, mark=o]
table [x=SNR_dB, y=OP_comp, col sep=comma] {figures/files_txt/plot2new/op_snr_fas_prp_5M_1K_5gamth_0.2kappa_1W.txt};
\addlegendentry{$P_{\text{out}}^{\text{p.r.p.}}$ (Eq. \eqref{eq76})};

\addplot[thick, black, dashed]
table [x=SNR_dB, y=OP_LB, col sep=comma] {figures/files_txt/plot2new/op_snr_fas_prp_5M_1K_5gamth_0.2kappa_1W.txt};
\addlegendentry{$P_{\text{out}}^{\text{p.r.p., LB}}$ (Eq. \eqref{eq107})};

\addplot[thick, black, densely dotted]
table [x=SNR_dB, y=OP_asymp, col sep=comma] {figures/files_txt/plot2new/op_snr_fas_prp_5M_1K_5gamth_0.2kappa_1W.txt};
\addlegendentry{$P_{\text{out} | \bar{\gamma} \gg 1 }^{\text{p.r.p.}}$ (Eq. \eqref{eq109})};

\addplot[thick, black, mark=+]
table [x=SNR_dB, y=OP_simul, col sep=comma] {figures/files_txt/plot2new/op_snr_fas_jakes_5M_1K_5gamth_0.2kappa_1W.txt};
\addlegendentry{$P_{\text{out}}$ (Jakes')};

\addplot[thick, black, mark=x, mark size=1.5pt]
table [x=SNR_dB, y=OP_simul, col sep=comma] {figures/files_txt/plot2new/op_snr_fas_rhom_5M_1K_5gamth_0.2kappa_1W.txt};
\addlegendentry{$P_{\text{out}}^{\text{p.r.p., dist.}}$};
\addplot[thick, blue, only marks, mark=asterisk, mark size=1.5pt]
table [x=SNR_dB, y=OP_simul, col sep=comma] {figures/files_txt/plot2new/op_snr_fas_prp_5M_4K_5gamth_0.2kappa_1W.txt};

\addplot[thick, blue, mark=o]
table [x=SNR_dB, y=OP_comp, col sep=comma] {figures/files_txt/plot2new/op_snr_fas_prp_5M_4K_5gamth_0.2kappa_1W.txt};

\addplot[thick, blue, dashed]
table [x=SNR_dB, y=OP_LB, col sep=comma] {figures/files_txt/plot2new/op_snr_fas_prp_5M_4K_5gamth_0.2kappa_1W.txt};

\addplot[thick, blue, densely dotted]
table [x=SNR_dB, y=OP_asymp, col sep=comma] {figures/files_txt/plot2new/op_snr_fas_prp_5M_4K_5gamth_0.2kappa_1W.txt};

\addplot[thick, blue, mark=+]
table [x=SNR_dB, y=OP_simul, col sep=comma] {figures/files_txt/plot2new/op_snr_fas_jakes_5M_4K_5gamth_0.2kappa_1W.txt};

\addplot[thick, blue, mark=x, mark size=1.5pt]
table [x=SNR_dB, y=OP_simul, col sep=comma] {figures/files_txt/plot2new/op_snr_fas_rhom_5M_4K_5gamth_0.2kappa_1W.txt};
\addplot[thick, red, only marks, mark=asterisk, mark size=1.5pt]
table [x=SNR_dB, y=OP_simul, col sep=comma] {figures/files_txt/plot2new/op_snr_fas_prp_10M_2K_5gamth_0.2kappa_1W.txt};

\addplot[thick, red, mark=o]
table [x=SNR_dB, y=OP_comp, col sep=comma] {figures/files_txt/plot2new/op_snr_fas_prp_10M_2K_5gamth_0.2kappa_1W.txt};

\addplot[thick, red, dashed]
table [x=SNR_dB, y=OP_LB, col sep=comma] {figures/files_txt/plot2new/op_snr_fas_prp_10M_2K_5gamth_0.2kappa_1W.txt};

\addplot[thick, red, densely dotted]
table [x=SNR_dB, y=OP_asymp, col sep=comma] {figures/files_txt/plot2new/op_snr_fas_prp_10M_2K_5gamth_0.2kappa_1W.txt};

\addplot[thick,  green!50!black, only marks, mark=asterisk, mark size=1.5pt] 
table [x=SNR_dB, y=OP_simul, col sep=comma] {figures/files_txt/plot2new/op_snr_fas_prp_10M_10K_5gamth_0.2kappa_1W.txt};

\addplot[thick,  green!50!black, mark=o]
table [x=SNR_dB, y=OP_comp, col sep=comma] {figures/files_txt/plot2new/op_snr_fas_prp_10M_10K_5gamth_0.2kappa_1W.txt};

\addplot[thick,  green!50!black, dashed]
table [x=SNR_dB, y=OP_LB, col sep=comma] {figures/files_txt/plot2new/op_snr_fas_prp_10M_10K_5gamth_0.2kappa_1W.txt};

\addplot[thick,  green!50!black, densely dotted]
table [x=SNR_dB, y=OP_asymp, col sep=comma] {figures/files_txt/plot2new/op_snr_fas_prp_10M_10K_5gamth_0.2kappa_1W.txt};

\node[draw, thin, ellipse, minimum width=0.2cm, minimum height=0.6cm, inner sep=0pt] at (axis cs:12,7e-4) {};

\node[draw, thin, ellipse, minimum width=0.2cm, minimum height=0.65cm, inner sep=0pt] at (axis cs:10,1e-4) {};

\node[draw, thin, ellipse, minimum width=0.2cm, minimum height=0.65cm, inner sep=0pt] at (axis cs:11,3.5e-8) {};

\node[draw, thin, ellipse, minimum width=0.2cm, minimum height=0.35cm, inner sep=0pt] at (axis cs:10,1e-10) {};

\begin{scope}[>=latex]
\draw[->, thin] (12,4e-3) -- (12.25,3.5e-2);
\draw[->, thin] (9.8,1e-4) -- (3.7,1e-5);
\draw[->, thin] (10.8,3.5e-8) -- (4.1,1e-6);
\draw[->, thin] (9.8,1e-10) -- (9,9e-12);
\end{scope}
\node[black, font=\scriptsize] at (13,7e-2) {$M = 5, K=1$};

\node[black, font=\scriptsize] at (1.8,1e-5) {$M=5, K=4$};

\node[black, font=\scriptsize] at (2,1e-6) {$M=10, K=2$};

\node[black, font=\scriptsize] at (8.5,3e-12) {$M=10, K=10$};
\end{semilogyaxis}

\end{tikzpicture}
\caption{\textcolor{black}{OP versus average SNR $\bar{\gamma}$ considering the physical reference port model, with $\gamma_{\text{th}}=5$~dB, $\kappa=0.2$, $W=1$, $M=\{5,10\}$, and $K=\{1,2,4,10\}$.}}
\label{f4}
\end{figure}

Figs.~\ref{f2} and~\ref{f5} plot the OP versus the number of selected ports $K$ considering the virtual and physical reference port models, respectively, for different values of $W$ and $M$, $\gamma_{\text{th}}=1$~dB, $\kappa=0.2$, and $\bar{\gamma}=5$~dB. The analytical expressions in \eqref{eq46} and \eqref{eq76} are illustrated together with their lower bound in \eqref{eq95} and \eqref{eq107} for the virtual and physical reference port models, respectively.
As expected, the OP reduces as $K$ increases due to the increased gain obtained from combining a larger number of favorable ports. Furthermore, the OP decreases as the spatial correlation among the ports decreases, which occurs for higher values of $W$. We also observe that the lower bound is tighter for the virtual reference port model than for the physical reference port model. The figures also present results for random port selection (labeled `R.P.S.'), where the $K$ ports are chosen randomly.\footnote{Note that the corresponding results do not depend on $M$.} Clearly, this strategy incurs a significant performance loss compared with selecting the best $K$ ports. \textcolor{black}{In addition, the figures include the OP of a $K$-branch MRC scheme subject to i.i.d. Rician fading channels. It is observed that the FAS with either reference port correlation model outperforms the $K$-branch MRC scheme, and this can always be ensured for sufficiently large $M$.}

\begin{figure}[!t]
\centering
\begin{tikzpicture}

\begin{semilogyaxis}[
width=3.4in, height=2.5in,
xmin=1, xmax=6,
ymin=1e-12, ymax=1e0,
xlabel={Selected ports $K$},
ylabel={Outage probability},
xlabel near ticks,
ylabel near ticks,
label style={font=\footnotesize},
title={$\gamma_{\text{th}}=1$ dB, $\kappa=0.2$, $\bar{\gamma}=5$ dB},
title style={font=\scriptsize, yshift=-2mm},
xtick={1,2,...,6},
ytick={1e-12,1e-10,1e-8,1e-6,1e-4,1e-2,1e0},
ticklabel style={font=\footnotesize},
legend style={at={(0.01,0.01)}, anchor=south west, font=\tiny, legend columns=2, inner sep=1pt, row sep=-3pt, fill opacity=0.6, draw opacity=1, text opacity=1},
legend cell align=left,
grid=major
]

\addplot[thick, blue, mark=o]
table [x=K_vals, y=OP_comp, col sep=comma] {figures/files_txt/plot3new/op_K_fas_vrp_6M_5gam_1gamth_0.2kappa_1W.txt};
\addlegendentry{$P_{\text{out}}^{\text{v.r.p.}}$, $W=1$};

\addplot[thick, blue, dashed, mark=o,mark options={solid}]
table [x=K_vals, y=OP_LB, col sep=comma] {figures/files_txt/plot3new/op_K_fas_vrp_6M_5gam_1gamth_0.2kappa_1W.txt};
\addlegendentry{${P}_{\text{out}}^{\text{v.r.p., LB}}$, $W=1$};

\addplot[thick, blue, mark=o]
table [x=K_vals, y=OP_comp, col sep=comma, forget plot] {figures/files_txt/plot3new/op_K_fas_vrp_10M_5gam_1gamth_0.2kappa_1W.txt};

\addplot[thick, blue, dashed, mark=o,mark options={solid}, forget plot]
table [x=K_vals, y=OP_LB, col sep=comma] {figures/files_txt/plot3new/op_K_fas_vrp_10M_5gam_1gamth_0.2kappa_1W.txt};

\addplot[thick, red, mark=asterisk]
table [x=K_vals, y=OP_comp, col sep=comma] {figures/files_txt/plot3new/op_K_fas_vrp_6M_5gam_1gamth_0.2kappa_2W.txt};
\addlegendentry{$P_{\text{out}}^{\text{v.r.p.}}$, $W=2$};

\addplot[thick, red, dashed, mark=asterisk, mark options={solid}]
table [x=K_vals, y=OP_LB, col sep=comma] {figures/files_txt/plot3new/op_K_fas_vrp_6M_5gam_1gamth_0.2kappa_2W.txt};
\addlegendentry{${P}_{\text{out}}^{\text{v.r.p., LB}}$, $W=2$};

\addplot[thick, red, mark=asterisk, forget plot]
table [x=K_vals, y=OP_comp, col sep=comma] {figures/files_txt/plot3new/op_K_fas_vrp_10M_5gam_1gamth_0.2kappa_2W.txt};

\addplot[thick, red, dashed, mark=asterisk, mark options={solid}, forget plot]
table [x=K_vals, y=OP_LB, col sep=comma] {figures/files_txt/plot3new/op_K_fas_vrp_10M_5gam_1gamth_0.2kappa_2W.txt};

\addplot[thick, black, mark=+]
table [x=K_vals, y=OP_simul, col sep=comma] {figures/files_txt/plot3new/op_K_fas_vrp_rps_5gam_1gamth_0.2kappa_1W.txt};
\addlegendentry{R.P.S. with MRC, $W=1$};

\addplot[thick, black, dashed, mark=+, mark options={solid}]
table [x=K_vals, y=OP_simul, col sep=comma] {figures/files_txt/plot3new/op_K_fas_vrp_rps_5gam_1gamth_0.2kappa_2W.txt};
\addlegendentry{R.P.S. with MRC, $W=2$};

\addplot[thick, black, densely dotted, mark=x, mark options={solid}]
table [x=K_vals, y=OP_simul, col sep=comma] {figures/files_txt/plot3new/op_K_mrc_5gam_1gamth_0.2kappa_2W.txt};
\addlegendentry{$K$-branch MRC};

\node[draw, thin, ellipse, minimum width=0.2cm, minimum height=0.5cm, inner sep=0pt] at (axis cs:3,4e-5) {};

\node[draw, thin, ellipse, minimum width=0.2cm, minimum height=0.5cm, inner sep=0pt] at (axis cs:5.5,8e-10) {};

\begin{scope}[>=latex]
\draw[->, thin] (3,9.5e-6) -- (3.3,8e-7);
\draw[->, thin] (5.5,3.4e-9) -- (5.5,1.3e-7);
\end{scope}
\node[black, font=\scriptsize] at (3.5,3e-7) {$M = 6$};

\node[black, font=\scriptsize] at (5.5,3e-7) {$M=10$};
\end{semilogyaxis}

\end{tikzpicture}
\caption{\textcolor{black}{OP versus number of selected ports $K$ considering the virtual reference port model, with $\gamma_{\text{th}}=1$~dB, $\kappa=0.2$, $W \in \{1, 2\}$, $M \in \{6,10\}$, and $\bar{\gamma}=5$~dB.}}
\label{f2}
\end{figure}

\begin{figure}[!t]
\centering
\begin{tikzpicture}

\begin{semilogyaxis}[
width=3.4in, height=2.5in,
xmin=1, xmax=6,
ymin=1e-12, ymax=1e0,
xlabel={Selected ports $K$},
ylabel={Outage probability},
xlabel near ticks,
ylabel near ticks,
label style={font=\footnotesize},
title={$\gamma_{\text{th}}=1$ dB, $\kappa=0.2$, $\bar{\gamma}=5$ dB},
title style={font=\scriptsize, yshift=-2mm},
xtick={1,2,...,6},
ytick={1e-12,1e-10,1e-8,1e-6,1e-4,1e-2,1e0},
ticklabel style={font=\footnotesize},
legend style={at={(0.01,0.01)}, anchor=south west, font=\tiny, legend columns=2, inner sep=1pt, row sep=-3pt, fill opacity=0.6, draw opacity=1, text opacity=1},
legend cell align=left,
grid=major
]

\addplot[thick, blue, mark=o]
table [x=K_vals, y=OP_comp, col sep=comma] {figures/files_txt/plot4new/op_K_fas_prp_6M_5gam_1gamth_0.2kappa_1W.txt};
\addlegendentry{$P_{\text{out}}^{\text{p.r.p}}$, $W=1$};

\addplot[thick, blue, dashed, mark=o,mark options={solid}]
table [x=K_vals, y=OP_LB, col sep=comma] {figures/files_txt/plot4new/op_K_fas_prp_6M_5gam_1gamth_0.2kappa_1W.txt};
\addlegendentry{${P}_{\text{out}}^{\text{p.r.p, LB}}$, $W=1$};

\addplot[thick, blue, mark=o]
table [x=K_vals, y=OP_comp, col sep=comma, forget plot] {figures/files_txt/plot4new/op_K_fas_prp_10M_5gam_1gamth_0.2kappa_1W.txt};

\addplot[thick, blue, dashed, mark=o,mark options={solid}, forget plot]
table [x=K_vals, y=OP_LB, col sep=comma] {figures/files_txt/plot4new/op_K_fas_prp_10M_5gam_1gamth_0.2kappa_1W.txt};

\addplot[thick, red, mark=asterisk]
table [x=K_vals, y=OP_comp, col sep=comma] {figures/files_txt/plot4new/op_K_fas_prp_6M_5gam_1gamth_0.2kappa_2W.txt};
\addlegendentry{$P_{\text{out}}^{\text{p.r.p}}$, $W=2$};

\addplot[thick, red, dashed, mark=asterisk, mark options={solid}]
table [x=K_vals, y=OP_LB, col sep=comma] {figures/files_txt/plot4new/op_K_fas_prp_6M_5gam_1gamth_0.2kappa_2W.txt};
\addlegendentry{${P}_{\text{out}}^{\text{p.r.p, LB}}$, $W=2$};

\addplot[thick, red, mark=asterisk, forget plot]
table [x=K_vals, y=OP_comp, col sep=comma] {figures/files_txt/plot4new/op_K_fas_prp_10M_5gam_1gamth_0.2kappa_2W.txt};

\addplot[thick, red, dashed, mark=asterisk, mark options={solid}, forget plot]
table [x=K_vals, y=OP_LB, col sep=comma] {figures/files_txt/plot4new/op_K_fas_prp_10M_5gam_1gamth_0.2kappa_2W.txt};

\addplot[thick, black, mark=+]
table [x=K_vals, y=OP_simul, col sep=comma] {figures/files_txt/plot4new/op_K_fas_prp_rps_5gam_1gamth_0.2kappa_1W.txt};
\addlegendentry{R.P.S. with MRC, $W=1$};

\addplot[thick, black, dashed, mark=+, mark options={solid}]
table [x=K_vals, y=OP_simul, col sep=comma] {figures/files_txt/plot4new/op_K_fas_prp_rps_5gam_1gamth_0.2kappa_2W.txt};
\addlegendentry{R.P.S. with MRC, $W=2$};

\addplot[thick, black, densely dotted, mark=x, mark options={solid}]
table [x=K_vals, y=OP_simul, col sep=comma] {figures/files_txt/plot4new/op_K_mrc_5gam_1gamth_0.2kappa_2W.txt};
\addlegendentry{$K$-branch MRC};

\node[draw, thin, ellipse, minimum width=0.2cm, minimum height=0.55cm, inner sep=0pt] at (axis cs:3,6e-5) {};

\node[draw, thin, ellipse, minimum width=0.2cm, minimum height=0.55cm, inner sep=0pt] at (axis cs:5.5,1.4e-9) {};

\begin{scope}[>=latex]
\draw[->, thin] (3,1.2e-5) -- (3.3,8e-7);
\draw[->, thin] (5.5,7e-9) -- (5.5,1.3e-7);
\end{scope}
\node[black, font=\scriptsize] at (3.5,3e-7) {$M = 6$};

\node[black, font=\scriptsize] at (5.5,3e-7) {$M=10$};
\end{semilogyaxis}

\end{tikzpicture}
\caption{\textcolor{black}{OP versus number of selected ports $K$ considering the physical reference port model, with $\gamma_{\text{th}}=1$~dB, $\kappa=0.2$, $W \in \{1, 2\}$, $M \in \{6,10\}$, and $\bar{\gamma}=5$~dB.}}
\label{f5}
\end{figure}

To quantify the performance difference between the virtual and physical reference port models, Fig.~\ref{f7} shows the ratio of the OPs $P_{\text{out}}^{\text{p.r.p.}}/P_{\text{out}}^{\text{v.r.p.}}$ versus the average SNR $\bar{\gamma}$, for different values of $M$ and $K$, $\gamma_{\text{th}}=5$~dB, $\kappa=0.2$, and $W=0.4$. Note that a ratio close to one indicates similar performance by the two models, while a ratio substantially greater than one implies that the physical reference port model diverges from the virtual reference port model. In the low-SNR region, the figure shows that the ratio remains close to one because all ports experience weak channel conditions and the AWGN term dominates. Consequently, the benefit of selecting the best $K$ ports becomes negligible, and the choice of the reference port in the system model and analytical framework results in similar outage performance. However, as $\bar{\gamma}$ grows, the ratio of the OPs increases monotonically before approaching the asymptotic value $C^{\rm p.r.p.}/C^{\rm v.r.p.}$, as evident from \eqref{eq97}--\eqref{eq110}. Furthermore, for a given $M$, the ratio converges to the same value at high average SNR regardless of the choice of $K$. Thus, the two models predict substantially different outage probabilities beyond the low-SNR region, with a persistent relative gap at high SNR.

\begin{figure}[!t]
\centering
\begin{tikzpicture}

\begin{semilogyaxis}[
width=3.4in, height=2.5in,
xmin=5, xmax=40,
ymin=5e-1, ymax=4e1,
xlabel={Average SNR $\bar{\gamma}$ (dB)},
ylabel={$P_{\text{out}}^{\text{p.r.p.}}/P_{\text{out}}^{\text{v.r.p.}}$},
xlabel near ticks,
ylabel near ticks,
label style={font=\footnotesize},
xtick={5,10,...,40},
ytick={1e2,1e1,1e0},
ticklabel style={font=\footnotesize},
title={$\gamma_{\text{th}}=5$ dB, $\kappa=0.2$, $W=0.4$},
title style={font=\scriptsize, yshift=-2mm},
legend style={at={(0.99,0.01)}, anchor=south east, font=\tiny, legend columns=1, inner sep=1pt, row sep=-3pt, fill opacity=0.6, draw opacity=1, text opacity=1},
legend cell align=left,
grid=major
]

\addplot[thick, black, mark=o]
table [x=SNR_dB, y=OP_ratio, col sep=comma] {figures/files_txt/plot7/op_snr_ratio_prp_vrp_10M_1K_5gamth_0.2kappa_0.4W.txt};
\addlegendentry{$M=10, K=1$};

\addplot[thick, blue, mark=square]
table [x=SNR_dB, y=OP_ratio, col sep=comma] {figures/files_txt/plot7/op_snr_ratio_prp_vrp_10M_5K_5gamth_0.2kappa_0.4W.txt};
\addlegendentry{$M=10, K=5$};

\addplot[thick, red, mark=asterisk]
table [x=SNR_dB, y=OP_ratio, col sep=comma] {figures/files_txt/plot7/op_snr_ratio_prp_vrp_10M_10K_5gamth_0.2kappa_0.4W.txt};
\addlegendentry{$M=10, K=10$};

\addplot[thick, black, dashed, mark=o, mark options={solid}]
table [x=SNR_dB, y=OP_ratio, col sep=comma] {figures/files_txt/plot7/op_snr_ratio_prp_vrp_15M_1K_5gamth_0.2kappa_0.4W.txt};
\addlegendentry{$M=15, K=1$};

\addplot[thick, blue, dashed, mark=square, mark options={solid}]
table [x=SNR_dB, y=OP_ratio, col sep=comma] {figures/files_txt/plot7/op_snr_ratio_prp_vrp_15M_5K_5gamth_0.2kappa_0.4W.txt};
\addlegendentry{$M=15, K=5$};

\addplot[thick, red, dashed, mark=asterisk, mark options={solid}]
table [x=SNR_dB, y=OP_ratio, col sep=comma] {figures/files_txt/plot7/op_snr_ratio_prp_vrp_15M_10K_5gamth_0.2kappa_0.4W.txt};
\addlegendentry{$M=15, K=10$};

\addplot[thick, black, dash dot, mark=o, mark options={solid}]
table [x=SNR_dB, y=OP_ratio, col sep=comma] {figures/files_txt/plot7/op_snr_ratio_prp_vrp_30M_1K_5gamth_0.2kappa_0.4W.txt};
\addlegendentry{$M=30, K=1$};

\addplot[thick, blue, dash dot, mark=square, mark options={solid}]
table [x=SNR_dB, y=OP_ratio, col sep=comma] {figures/files_txt/plot7/op_snr_ratio_prp_vrp_30M_5K_5gamth_0.2kappa_0.4W.txt};
\addlegendentry{$M=30, K=5$};

\addplot[thick, red, dash dot, mark=asterisk, mark options={solid}]
table [x=SNR_dB, y=OP_ratio, col sep=comma] {figures/files_txt/plot7/op_snr_ratio_prp_vrp_30M_10K_5gamth_0.2kappa_0.4W.txt};
\addlegendentry{$M=30, K=10$};

\end{semilogyaxis}

\end{tikzpicture} \vspace{5mm}
\caption{\textcolor{black}{Ratio of the OPs $P_{\text{out}}^{\text{p.r.p.}}/P_{\text{out}}^{\text{v.r.p.}}$ versus average SNR $\bar{\gamma}$, with $\gamma_{\text{th}}=5$~dB, $\kappa=0.2$, $W=0.4$, $M \in \{ 10, 15, 30\}$, and $K \in \{1,5,10 \}$.}}
\label{f7}
\end{figure}

\begin{figure}[!t]
\centering
\begin{tikzpicture}

\begin{axis}[
width=3.4in, height=2.5in,
xmin=1, xmax=10,
ymin=-6, ymax=15,
xlabel={Selected ports $K$},
ylabel={$P_{\text{out}}^{\text{p.r.p.}}/P_{\text{out}}^{\text{v.r.p.}}$},
xlabel near ticks,
ylabel near ticks,
label style={font=\footnotesize},
title={$\gamma_{\text{th}}=1$ dB, $\kappa=0.2$, $\bar{\gamma}=15$ dB},
title style={font=\scriptsize, yshift=-2mm},
xtick={1,2,...,10},
ytick={15,12,...,-3},
ticklabel style={font=\footnotesize},
legend style={at={(0.01,0.01)}, anchor=south west, font=\tiny, legend columns=2, inner sep=1pt, row sep=-3pt, fill opacity=0.6, draw opacity=1, text opacity=1},
legend cell align=left,
grid=major
]

\addplot[thick, blue, mark=triangle]
table [x=Kvals, y=OP_ratio, col sep=comma] {figures/files_txt/plot8/op_K_ratio_prp_vrp_10M_15gam_1gamth_0.2kappa_0.5W.txt};
\addlegendentry{$M=10, W=0.5$};

\addplot[thick, blue, dashed, mark=triangle, mark options={solid}]
table [x=Kvals, y=OP_ratio, col sep=comma]
{figures/files_txt/plot8/op_K_ratio_prp_vrp_10M_15gam_1gamth_0.2kappa_2W.txt};
\addlegendentry{$M=10, W=2$};

\addplot[thick, blue, dash dot, mark=triangle, mark options={solid}]
table [x=Kvals, y=OP_ratio, col sep=comma]
{figures/files_txt/plot8/op_K_ratio_prp_vrp_10M_15gam_1gamth_0.2kappa_5W.txt};
\addlegendentry{$M=10, W=5$};

\addplot[thick, red, mark=square]
table [x=Kvals, y=OP_ratio, col sep=comma]
{figures/files_txt/plot8/op_K_ratio_prp_vrp_15M_15gam_1gamth_0.2kappa_0.5W.txt};
\addlegendentry{$M=15, W=0.5$};

\addplot[thick, red, dashed, mark=square, mark options={solid}]
table [x=Kvals, y=OP_ratio, col sep=comma]
{figures/files_txt/plot8/op_K_ratio_prp_vrp_15M_15gam_1gamth_0.2kappa_2W.txt};
\addlegendentry{$M=15, W=2$};

\addplot[thick, red, dash dot, mark=square, mark options={solid}]
table [x=Kvals, y=OP_ratio, col sep=comma]
{figures/files_txt/plot8/op_K_ratio_prp_vrp_15M_15gam_1gamth_0.2kappa_5W.txt};
\addlegendentry{$M=15, W=5$};
\addplot[thick, black, mark=*]
table [x=Kvals, y=OP_ratio, col sep=comma]
{figures/files_txt/plot8/op_K_ratio_prp_vrp_30M_15gam_1gamth_0.2kappa_0.5W.txt};
\addlegendentry{$M=30, W=0.5$};

\addplot[thick, black, dashed, mark=*, mark options={solid}]
table [x=Kvals, y=OP_ratio, col sep=comma]
{figures/files_txt/plot8/op_K_ratio_prp_vrp_30M_15gam_1gamth_0.2kappa_2W.txt};
\addlegendentry{$M=30, W=2$};

\addplot[thick, black, dash dot, mark=*, mark options={solid}]
table [x=Kvals, y=OP_ratio, col sep=comma]
{figures/files_txt/plot8/op_K_ratio_prp_vrp_30M_15gam_1gamth_0.2kappa_5W.txt};
\addlegendentry{$M=30, W=5$};

\end{axis}

\end{tikzpicture} \vspace{5mm}
\caption{\textcolor{black}{Ratio of the OPs $P_{\text{out}}^{\text{p.r.p.}}/P_{\text{out}}^{\text{v.r.p.}}$ versus number of selected ports $K$, with $\gamma_{\text{th}}=1$~dB, $\kappa=0.2$, $W \in \{0.5, 2, 5\}$, $M \in \{10,15,30\}$, and $\bar{\gamma}=15$~dB.}}
\label{f8}
\end{figure}

A similar comparison is shown in Fig.~\ref{f8}, which plots the ratio of the OPs $P_{\text{out}}^{\text{p.r.p.}}/P_{\text{out}}^{\text{v.r.p.}}$ versus the number of selected ports $K$ for different values of $W$ and $M$, $\gamma_{\text{th}}=1$~dB, $\kappa=0.2$, and $\bar{\gamma}=15$~dB. It is observed that the ratio of the OPs using the two models with an increasing number of selected ports remains nearly the same for small values of $M$. However, as $M$ increases, the ratio initially increases with $K$ before saturating. This implies that the performance difference arising from choosing the physical reference port model becomes significant with an increase in the number of selected ports, when the FAS employs a higher number of ports with small inter-port distances.

\begin{figure}[!t]
\centering
\begin{tikzpicture}

\begin{semilogyaxis}[
	width=3.4in, height=2.5in,
	xmin=0, xmax=10,
	ymin=1e-7, ymax=1e0,
	xlabel={Rician factor $\kappa$},
	ylabel={$P_{\text{out}}^{\text{v.r.p.}}$},
    title={$\gamma_{\text{th}}=1$ dB, $W=0.5$, $\bar{\gamma}=5$ dB},
	xlabel near ticks,
	ylabel near ticks,
	label style={font=\footnotesize},
	xtick={0,1,2,...,10},
    ytick={1e-7,1e-6,1e-5,1e-4,1e-3,1e-2,1e-1,1e0},
    ticklabel style={font=\footnotesize},
	legend style={at={(0.99,0.01)}, anchor=south east, font=\tiny, legend columns=2, inner sep=1pt, row sep=-3pt, fill opacity=0.6, draw opacity=1, text opacity=1},
	legend cell align=left,
	title style={font=\scriptsize, yshift=-2mm},
	grid=major
]

\addplot[thick, blue, mark=asterisk]
table [x=kappa_vec, y=OP_comp, col sep=comma] {figures/files_txt/plot7new/op_kappa_fas_vrp_5M_2K_1gamth_5gam_0.5W.txt};
\addlegendentry{$M=5$, $K=2$};

\addplot[thick, blue, mark=o]
table [x=kappa_vec, y=OP_comp, col sep=comma] {figures/files_txt/plot7new/op_kappa_fas_vrp_5M_4K_1gamth_5gam_0.5W.txt};
\addlegendentry{$M=5$, $K=4$};

\addplot[thick, blue, mark=square]
table [x=kappa_vec, y=OP_comp, col sep=comma] {figures/files_txt/plot7new/op_kappa_fas_vrp_5M_5K_1gamth_5gam_0.5W.txt};
\addlegendentry{$M=5$, $K=5$};

\addplot[thick, red, dashed, mark=asterisk, mark options={solid}]
table [x=kappa_vec, y=OP_comp, col sep=comma] {figures/files_txt/plot7new/op_kappa_fas_vrp_10M_2K_1gamth_5gam_0.5W.txt};
\addlegendentry{$M=10$, $K=2$};

\addplot[thick, red, dashed, mark=o, mark options={solid}]
table [x=kappa_vec, y=OP_comp, col sep=comma] {figures/files_txt/plot7new/op_kappa_fas_vrp_10M_4K_1gamth_5gam_0.5W.txt};
\addlegendentry{$M=10$, $K=4$};

\addplot[thick, red, dashed, mark=square, mark options={solid}]
table [x=kappa_vec, y=OP_comp, col sep=comma] {figures/files_txt/plot7new/op_kappa_fas_vrp_10M_10K_1gamth_5gam_0.5W.txt};
\addlegendentry{$M=10$, $K=10$};

\addplot[thick, black, mark=*, mark options={solid}]
table [x=kappa_vec, y=OP_comp, col sep=comma] {figures/files_txt/plot7new/op_kappa_fas_vrp_rps_5M_2K_1gamth_5gam_0.5W.txt};
\addlegendentry{R.P.S. with MRC, $K=2$};

\addplot[thick, black, dashed, mark=*, mark options={solid}]
table [x=kappa_vec, y=OP_comp, col sep=comma] {figures/files_txt/plot7new/op_kappa_fas_vrp_rps_5M_4K_1gamth_5gam_0.5W.txt};
\addlegendentry{R.P.S. with MRC, $K=4$};

\addplot[thick, green!50!black, mark=x, mark options={solid}]
table [x=kappa_vec, y=OP_comp, col sep=comma] {figures/files_txt/plot7new/op_kappa_mrc_2K_1gamth_5gam.txt};
\addlegendentry{$K=2$-branch MRC};

\addplot[thick, green!50!black, dashed, mark=x, mark options={solid}]
table [x=kappa_vec, y=OP_comp, col sep=comma] {figures/files_txt/plot7new/op_kappa_mrc_4K_1gamth_5gam.txt};
\addlegendentry{$K=4$-branch MRC};

\end{semilogyaxis}

\end{tikzpicture}
\caption{\textcolor{black}{$P_{\text{out}}^{\text{v.r.p.}}$ (with best $K$ ports and random port selection) versus Rician factor $\kappa$ considering the virtual reference port model, with $\gamma_{\text{th}}=1$~dB, $W=0.5$, $M \in \{5,10\}$, $K \in \{2,4,M\}$, and $\bar{\gamma}=5$~dB.}}
\label{f3}
\end{figure}

\begin{figure}[!t]
\centering
\begin{tikzpicture}

\begin{semilogyaxis}[
	width=3.4in, height=2.5in,
	xmin=0, xmax=10,
	ymin=1e-7, ymax=1e0,
	xlabel={Rician factor $\kappa$},
	ylabel={$P_{\text{out}}^{\text{p.r.p.}}$},
    title={$\gamma_{\text{th}}=1$ dB, $W=0.5$, $\bar{\gamma}=5$ dB},
	xlabel near ticks,
	ylabel near ticks,
	label style={font=\footnotesize},
	xtick={0,1,2,...,10},
    ytick={1e-7,1e-6,1e-5,1e-4,1e-3,1e-2,1e-1,1e0},
    ticklabel style={font=\footnotesize},
	legend style={at={(0.99,0.01)}, anchor=south east, font=\tiny, legend columns=2, inner sep=1pt, row sep=-3pt, fill opacity=0.6, draw opacity=1, text opacity=1},
	legend cell align=left,
	title style={font=\scriptsize, yshift=-2mm},
	grid=major
]

\addplot[thick, blue, mark=asterisk]
table [x=kappa_vec, y=OP_comp, col sep=comma] {figures/files_txt/plot8new/op_kappa_fas_prp_5M_2K_1gamth_5gam_0.5W.txt};
\addlegendentry{$M=5$, $K=2$};

\addplot[thick, blue, mark=o]
table [x=kappa_vec, y=OP_comp, col sep=comma] {figures/files_txt/plot8new/op_kappa_fas_prp_5M_4K_1gamth_5gam_0.5W.txt};
\addlegendentry{$M=5$, $K=4$};

\addplot[thick, blue, mark=square]
table [x=kappa_vec, y=OP_comp, col sep=comma] {figures/files_txt/plot8new/op_kappa_fas_prp_5M_5K_1gamth_5gam_0.5W.txt};
\addlegendentry{$M=5$, $K=5$};

\addplot[thick, red, dashed, mark=asterisk, mark options={solid}]
table [x=kappa_vec, y=OP_comp, col sep=comma] {figures/files_txt/plot8new/op_kappa_fas_prp_10M_2K_1gamth_5gam_0.5W.txt};
\addlegendentry{$M=10$, $K=2$};

\addplot[thick, red, dashed, mark=o, mark options={solid}]
table [x=kappa_vec, y=OP_comp, col sep=comma] {figures/files_txt/plot8new/op_kappa_fas_prp_10M_4K_1gamth_5gam_0.5W.txt};
\addlegendentry{$M=10$, $K=4$};

\addplot[thick, red, dashed, mark=square, mark options={solid}]
table [x=kappa_vec, y=OP_comp, col sep=comma] {figures/files_txt/plot8new/op_kappa_fas_prp_10M_10K_1gamth_5gam_0.5W.txt};
\addlegendentry{$M=10$, $K=10$};

\addplot[thick, black, mark=*, mark options={solid}]
table [x=kappa_vec, y=OP_comp, col sep=comma] {figures/files_txt/plot8new/op_kappa_fas_prp_rps_5M_2K_1gamth_5gam_0.5W.txt};
\addlegendentry{R.P.S. with MRC, $K=2$};

\addplot[thick, black, dashed, mark=*, mark options={solid}]
table [x=kappa_vec, y=OP_comp, col sep=comma] {figures/files_txt/plot8new/op_kappa_fas_prp_rps_5M_4K_1gamth_5gam_0.5W.txt};
\addlegendentry{R.P.S. with MRC, $K=4$};

\addplot[thick, green!50!black, mark=x, mark options={solid}]
table [x=kappa_vec, y=OP_comp, col sep=comma] {figures/files_txt/plot8new/op_kappa_mrc_2K_1gamth_5gam.txt};
\addlegendentry{$K=2$-branch MRC};

\addplot[thick, green!50!black, dashed, mark=x, mark options={solid}]
table [x=kappa_vec, y=OP_comp, col sep=comma] {figures/files_txt/plot8new/op_kappa_mrc_4K_1gamth_5gam.txt};
\addlegendentry{$K=4$-branch MRC};
\end{semilogyaxis}

\end{tikzpicture}
\caption{\textcolor{black}{$P_{\text{out}}^{\text{p.r.p.}}$ (with best $K$ ports and random port selection) versus Rician factor $\kappa$ considering the physical reference port model, with $\gamma_{\text{th}}=1$~dB, $W=0.5$, $M \in \{5,10\}$, $K \in \{2,4,M\}$, and $\bar{\gamma}=5$~dB.}}
\label{f6}
\end{figure}

Lastly, Figs.~\ref{f3} and \ref{f6} show the analytical OP in \eqref{eq46} and \eqref{eq76} considering the virtual and physical reference port models, respectively, versus the Rician factor $\kappa$ for different values of $M$ and $K$, $\gamma_{\text{th}}=1$ dB, $W=0.5$, and $\bar{\gamma}=5$~dB. As expected, selecting the best $K$ ports significantly outperforms random port selection for MRC and the $K$-branch MRC scheme with a sufficiently large number of ports. Interestingly, it is observed that the OP is a concave function of $\kappa$, and this concavity becomes more pronounced as $M$ and $K$ increase, with the peak shifting towards larger values of $\kappa$. This implies that the performance gain due to the LoS component becomes more prominent as the Rician factor $\kappa$ increases only for larger values of $M$ and $K$. Beyond the maximum, the OP consistently reduces with the Rician factor, which is indicative of a superior performance for a stronger LoS component. \textcolor{black}{The non-monotonic trend reflects how $\kappa$ redistributes power between the LoS and scattered component at fixed $\bar{\gamma}$. The scattered component is the only part that varies across ports, and hence the only source of selection and combining gains, whereas the LoS component is common to all ports and merely lifts the received SNR. For small $\kappa$, the power of the scattered component in the $M-1$ diversity-carrying spatial modes shrinks by a factor of $1/(\kappa+1)$ without LoS compensation, making simultaneous deep fades more likely and thus increasing the OP. For large $\kappa$, the LoS component lifts the SNR well above $\gamma_{\text{th}}$, and an outage would require the scattered component to cancel it, which is exponentially unlikely and thus decreases the OP. At high SNR, this trade-off is captured by the $\kappa$-dependent factors in \eqref{eq97}--\eqref{eq110}, whose maximum lies at
\beq
\kappa^\star_{\text{v.r.p.}} = (M-1)\rho^2 , \qquad
\kappa^\star_{\text{p.r.p.}} = \frac{(M-1)(1-\tau)}{1+(M-1)\tau},
\label{eqKstar}
\eeq
respectively. For uncorrelated ports, we have $\kappa^\star=0$. Therefore, the non-monotonicity is caused by spatial correlation, which confines the common LoS component to a single spatial mode.}
\section{Conclusions} \label{sec:con}
We considered a FAS with a single-antenna transmitter communicating with a receiver equipped with a one-dimensional FA comprising $M$ ports over Rician fading channels. The receiver selects the best $K$ ports among the $M$ available FA ports based on their instantaneous SNR and combines them using MRC. To incorporate the port correlation into the channels, we considered both the virtual and physical reference port models. For each model, a CF-based approach was used to obtain the statistics of the post-combining SNR along with integral-form expressions for the OP. Additionally, a tight lower bound and the asymptotic OP of the FAS at high average SNR were derived, revealing the diversity order of the FAS to be equal to $M$. Numerical results were presented to validate the analytical framework and showed that the performance gap between the two models becomes more pronounced as the average SNR and the number of available and selected ports increase. Moreover, both models were shown to achieve the same diversity order as Jakes' benchmark model, while the performance of the virtual reference port model is closer to the distance-dependent physical reference port model. \textcolor{black}{Further directions for this research may include estimation errors due to imperfect CSI, which would jointly affect the selection and ranking of the best ports.}
\balance
\bibliographystyle{IEEEtran}
\bibliography{IEEEabrv,bibliography}

@ARTICLE{10753482,
  author={New, Wee Kiat and others},
  journal={IEEE Commun. Surveys Tuts.}, 
  title={\textcolor{black}{A tutorial on fluid antenna system for 6G networks: Encompassing communication theory, optimization methods and hardware designs}}, 
  year={2025},
  volume={27},
  number={4},
  pages={2325--2377},
  doi={10.1109/COMST.2024.3498855}}

@article{LIU2026103261,
title = {\textcolor{black}{Outage probability and ergodic capacity analysis of fluid antenna systems over {Fisher-Snedecor $F$} composite fading channels}},
journal = {Physical Commun.},
volume = {78},
pages = {103261 (1--14)},
year = {2026},
issn = {1874-4907},
doi = {https://doi.org/10.1016/j.phycom.2026.103261},
author = {Zhedong Liu and Hui Zhao and Hongjiang Lei},
}

@ARTICLE{11646997,
  author={Yao, Jin and others},
  journal={IEEE Trans. Veh. Technol.}, 
  title={\textcolor{black}{Fluid antenna systems with terrestrial interference for countering UAV covert communications}}, 
  year={2026},
  volume={early Access},
  number={},
  pages={},
  doi={10.1109/TVT.2026.3722371}}

@ARTICLE{11535420,
  author={Bataineh, Mohammad Al and Albataineh, Zaid},
  journal={IEEE Open J. Commun. Soc.}, 
  title={\textcolor{black}{Outage analysis and diversity order of RIS-assisted energy-harvesting decode-and-forward relaying over $\alpha$-$\eta$-$\mu$ generalized fading channels}}, 
  year={2026},
  volume={7},
  number={},
  pages={5757--5773},
  doi={10.1109/OJCOMS.2026.3696645}}

@ARTICLE{11535468,
  author={Bataineh, Mohammad Al and Albataineh, Zaid},
  journal={IEEE Open J. Commun. Soc.}, 
  title={\textcolor{black}{Vine copula-based outage analysis for fluid antenna RIS-NOMA systems over Nakagami-$m$ fading channels}}, 
  year={2026},
  volume={7},
  number={},
  pages={5689--5710},
  doi={10.1109/OJCOMS.2026.3696777}}

@ARTICLE{11558417,
  author={Upadhya, Abhijeet and others},
  journal={IEEE Open J. Commun. Soc.}, 
  title={\textcolor{black}{On the evaluation of fluid antenna system-assisted cooperative relaying with diversity combining}}, 
  year={2026},
  volume={7},
  number={},
  pages={6703--6722},
  doi={10.1109/OJCOMS.2026.3702511}}

@ARTICLE{10468625,
  author={Vega-Sánchez, José David and others},
  journal={IEEE Trans. Veh. Technol.}, 
  title={\textcolor{black}{Fluid antenna system: Secrecy outage probability analysis}}, 
  year={2024},
  volume={73},
  number={8},
  pages={11458--11469},
  doi={10.1109/TVT.2024.3376475}}

@article{dinis2026,
  author  = {D. Dinis and R. Wichman},
  title   = {\textcolor{black}{Spatio-temporal port selection in fluid antennas under switching delays}},
  journal = {IEEE Commun. Lett.},
  volume  = {30},
  pages = {992--996},
  year = {2026}
}

@ARTICLE{10740058,
  author={Zhang, Jichen and others},
  journal={IEEE Open J. Antennas Propag.}, 
  title={\textcolor{black}{A novel pixel-based reconfigurable antenna applied in fluid antenna systems with high switching speed}}, 
  year={2025},
  volume={6},
  number={1},
  pages={212--228},
  doi={10.1109/OJAP.2024.3489215}
}

@INPROCEEDINGS{ganeshgc25,
  author={Tummi Ganesh and Soumya P. Dash and George C. Alexandropoulos},
  booktitle={Proc. IEEE Global Commun. Conf. (GLOBECOM)}, 
  title={Outage Probability Analysis of {RIS}-Assisted Fluid Antenna Systems over Double-{Nakagami}-$m$ Fading Channels}, 
  year={2025},
  doi={10.1109/WCNC61545.2025.10978677}}

@misc{dash2026,
      title={Symbol Error Analysis for Fluid Antenna Systems with One- and Two-Dimensional Modulation Schemes}, 
      author={Soumya P. Dash and George C. Alexandropoulos},
      year={2026},
      eprint={2604.06852},
      archivePrefix={arXiv},
      primaryClass={eess.SP},
      url={https://arxiv.org/abs/2604.06852}, 
}

@INPROCEEDINGS{10279614,
  author={Tlebaldiyeva, Leila and Arzykulov, Sultangali and Rabie, Khaled M. and Li, Xingwang and Nauryzbayev, Galymzhan},
  booktitle={Proc. IEEE Int. Conf. Commun. (ICC)}, 
  title={Outage Performance of Fluid Antenna System ({FAS})-aided Terahertz Communication Networks}, 
  year={2023},
  volume={},
  number={},
  doi={10.1109/ICC45041.2023.10279614}}

@ARTICLE{10308583,
  author={Vega-Sánchez, José David and López-Ramírez, Arianna Estefanía and Urquiza-Aguiar, Luis and Osorio, Diana Pamela Moya},
  journal={IEEE Wireless Commun. Lett.}, 
  title={Novel Expressions for the Outage Probability and Diversity Gains in Fluid Antenna System}, 
  year={2024},
  volume={13},
  number={2},
  pages={372--376},
  doi={10.1109/LWC.2023.3329780}}

@ARTICLE{11106811,
  author={Pakravan, Saeid and Ahmadzadeh, Mohsen and Zeng, Ming and Yang, Zhaohui and Hodtani, Ghosheh Abed and Chouinard, Jean-Yves and Pham, Quoc-Viet},
  journal={IEEE Trans. Veh. Technol.}, 
  title={Fluid Antenna-Assisted Uplink {NOMA} Networks Under Imperfect {SIC}}, 
  year={2026},
  volume={75},
  number={1},
  pages={1689--1694},
  doi={10.1109/TVT.2025.3594998}}

@INPROCEEDINGS{11432102,
  author={Inwood, Amy S. and Smith, Peter J. and Senanayake, Rajitha and Matthaiou, Michail},
  booktitle={Proc. IEEE Global Commun. Conf. (GLOBECOM)}, 
  title={High {SNR} Probabilities of Continuous Fluid Antenna Systems in {Ricean} Environments}, 
  year={2025},
  volume={},
  number={},
  doi={10.1109/GLOBECOM59602.2025.11432102}}

@ARTICLE{10980171,
  author={Lin, Xiao and Zhao, Yizhe and Yang, Halvin and Hu, Jie and Wong, Kai-Kit},
  journal={IEEE Trans. Wireless Commun.}, 
  title={Fluid Antenna Multiple Access Assisted Integrated Data and Energy Transfer: Outage and Multiplexing Gain Analysis}, 
  year={2025},
  volume={24},
  number={9},
  pages={7777--7793},
  doi={10.1109/TWC.2025.3562921}}

@ARTICLE{10855346,
  author={Yao, Junteng and Zhou, Liaoshi and Wu, Tuo and Jin, Ming and Pan, Cunhua and Elkashlan, Maged and Wong, Kai-Kit},
  journal={IEEE Trans. Wireless Commun.}, 
  title={Exploring Fairness for {FAS}-Assisted Communication Systems: From {NOMA} to {OMA}}, 
  year={2025},
  volume={24},
  number={4},
  pages={3433--3449},
  doi={10.1109/TWC.2025.3531056}}

@ARTICLE{11185052,
  author={Chen, Yu and Xu, Bowen and Li, Shijie and Cui, Qimei and Tao, Xiaofeng},
  journal={IEEE J. Sel. Areas Commun.}, 
  title={Analysis and Optimization for Low-Latency Communications in Slow Fluid Antenna Multiple Access Systems}, 
  year={2026},
  volume={44},
  number={},
  pages={1290--1306},
  doi={10.1109/JSAC.2025.3616075}}

@ARTICLE{11184548,
  author={Yang, Halvin and Derakhshani, Mahsa and Lambotharan, Sangarapillai and Hanzo, Lajos},
  journal={IEEE J. Sel. Areas Commun.}, 
  title={Performance Analysis of Fluid Antenna System Aided {OTFS} Satellite Communications}, 
  year={2026},
  volume={44},
  number={},
  pages={1092--1109},
  doi={10.1109/JSAC.2025.3615566}}

@ARTICLE{11371611,
  author={Dinis, Daniel and Wichman, Risto},
  journal={IEEE Wireless Commun. Lett.}, 
  title={{s-FAMA-GP}: A Low-Complexity Slow {FAMA} Using Interference Interpolation}, 
  year={2026},
  volume={15},
  number={},
  pages={1727--1731},
  doi={10.1109/LWC.2026.3661418}}

@ARTICLE{10924151,
  author={Zhao, Hui and Slock, Dirk},
  journal={IEEE Wireless Commun. Lett.}, 
  title={Analytical Insights Into Outage Probability and Ergodic Capacity of Fluid Antenna Systems}, 
  year={2025},
  volume={14},
  number={5},
  pages={1581--1585},
  doi={10.1109/LWC.2025.3550820}}

@ARTICLE{11459144,
  author={Huang, Zhiyu and Li, Guyue and Xu, Hao and Ng, Derrick Wing Kwan},
  journal={IEEE J. Sel. Areas Commun.}, 
  title={Fluid Antenna System-Assisted Physical Layer Secret Key Generation}, 
  year={2026},
  volume={44},
  number={},
  pages={4429--4443},
  doi={10.1109/JSAC.2026.3679311}}

@ARTICLE{10188603,
  author={Psomas, Constantinos and Kraidy, Ghassan M. and Wong, Kai-Kit and Krikidis, Ioannis},
  journal={IEEE Trans. Wireless Commun.}, 
  title={On the Diversity and Coded Modulation Design of Fluid Antenna Systems}, 
  year={2024},
  volume={23},
  number={3},
  pages={2082--2096},
  doi={10.1109/TWC.2023.3294997}}

@ARTICLE{10858773,
  author={Yao, Junteng and Zheng, Jianchao and Wu, Tuo and Jin, Ming and Yuen, Chau and Wong, Kai-Kit and Adachi, Fumiyuki},
  journal={IEEE Trans. Veh. Technol.}, 
  title={{FAS-RIS} Communication: Model, Analysis, and Optimization}, 
  year={2025},
  volume={74},
  number={6},
  pages={9938--9943},
  doi={10.1109/TVT.2025.3537294}}

@ARTICLE{10103838,
  author={Khammassi, Malek and Kammoun, Abla and Alouini, Mohamed-Slim},
  journal={IEEE Trans. Wireless Commun.}, 
  title={A New Analytical Approximation of the Fluid Antenna System Channel}, 
  year={2023},
  volume={22},
  number={12},
  pages={8843--8858},
  doi={10.1109/TWC.2023.3266411}}

@ARTICLE{10379539,
  author={Z. Wang and others},
  journal={IEEE Commun. Surveys Tuts.},
  title={A Tutorial on Extremely Large-Scale {MIMO} for {6G}: Fundamentals, Signal Processing, and Applications},
  year={2024},
  volume={26},
  number={3},
  pages={1560--1605},
  doi={10.1109/COMST.2023.3349276}
}

@ARTICLE{9770295,
  author={A. Shojaeifard and others},
  journal={Proc. IEEE},
  title={{MIMO} Evolution Beyond {5G} Through Reconfigurable Intelligent Surfaces and Fluid Antenna Systems},
  year={2022},
  volume={110},
  number={9},
  pages={1244--1265},
  doi={10.1109/JPROC.2022.3170247}
}

@ARTICLE{9349624,
  author={W. Jiang and B. Han and M. A. Habibi and H. D. Schotten},
  journal={IEEE Open J. Commun. Soc.},
  title={The Road Towards {6G}: A Comprehensive Survey},
  year={2021},
  volume={2},
  pages={334--366},
  doi={10.1109/OJCOMS.2021.3057679}
}

@ARTICLE{constcorr,
  author={K.-K. Wong and K.-F. Tong and Y. Chen and Y. Zhang},
  journal={Electron. Lett.},
  title={Closed-Form Expressions for Spatial Correlation Parameters for Performance Analysis of Fluid Antenna Systems},
  year={2022},
  volume={58},
  number={11},
  pages={454--457},
  doi={10.1049/ell2.12487}
}

@ARTICLE{9264694,
  author={K.-K. Wong and A. Shojaeifard and K.-F. Tong and Y. Zhang},
  journal={IEEE Trans. Wireless Commun.},
  title={Fluid Antenna Systems},
  year={2021},
  volume={20},
  number={3},
  pages={1950--1962},
  doi={10.1109/TWC.2020.3037595}
}

@ARTICLE{10146286,
  author={K.-K. Wong and K.-F. Tong and C.-B. Chae},
  journal={IEEE Commun. Lett.},
  title={Fluid Antenna System—{Part} {II}: Research Opportunities},
  year={2023},
  volume={27},
  number={8},
  pages={1924--1928},
  doi={10.1109/LCOMM.2023.3284318}
}

@ARTICLE{11302793,
  author={T. Wu and K. Zhi and J. Yao and others},
  journal={IEEE Wireless Commun.},
  title={Fluid Antenna Systems Enabling {6G}: Principles, Applications, and Research Directions},
  year={2025},
  pages={1--9},
  doi={10.1109/MWC.2025.3629597}
}

@ARTICLE{11247926,
  author={W. K. New and K.-K. Wong and C. Wang and C.-B. Chae and R. Murch and H. Jafarkhani and Y. Hao},
  journal={IEEE J. Sel. Areas Commun.},
  title={Fluid Antenna Systems: Redefining Reconfigurable Wireless Communications},
  year={2026},
  volume={44},
  pages={1013--1044},
  doi={10.1109/JSAC.2025.3632097}
}

@INPROCEEDINGS{11093131,
  author={A. F. M. S. Shah and others},
  booktitle={Proc. IEEE IEMCON},
  title={Performance Analysis of Fluid Antenna Multiple Access {(FAMA)} for {6G}},
  year={2024},
  doi={10.1109/IEMCON62851.2024.11093131}
}

@ARTICLE{9131873,
  author={K.-K. Wong and A. Shojaeifard and K.-F. Tong and Y. Zhang},
  journal={IEEE Commun. Lett.},
  title={Performance Limits of Fluid Antenna Systems},
  year={2020},
  volume={24},
  number={11},
  pages={2469--2472},
  doi={10.1109/LCOMM.2020.3006554}
}

@ARTICLE{10078147,
  author={K.-K. Wong and K.-F. Tong and Y. Chen and Y. Zhang and C.-B. Chae},
  journal={IEEE Trans. Wireless Commun.},
  title={Opportunistic Fluid Antenna Multiple Access},
  year={2023},
  volume={22},
  number={11},
  pages={7819--7833},
  doi={10.1109/TWC.2023.3255940}
}

@ARTICLE{10066316,
  author={K.-K. Wong and D. Morales-Jimenez and K.-F. Tong and C.-B. Chae},
  journal={IEEE Trans. Commun.},
  title={Slow Fluid Antenna Multiple Access},
  year={2023},
  volume={71},
  number={5},
  pages={2831--2846},
  doi={10.1109/TCOMM.2023.3255904}
}

@INPROCEEDINGS{9833952,
  author={P. Mukherjee and C. Psomas and I. Krikidis},
  booktitle={Proc. IEEE Int. Workshop Signal Process. Adv. Wireless Commun. (SPAWC)},
  title={On the Level Crossing Rate of Fluid Antenna Systems},
  year={2022},
  doi={10.1109/SPAWC51304.2022.9833952}
}

@INPROCEEDINGS{10167904,
  author={L. Tlebaldiyeva and S. Arzykulov and T. A. Tsiftsis and G. Nauryzbayev},
  booktitle={Proc. Int. Balkan Conf. Commun. Netw. (BalkanCom)},
  title={Full-Duplex Cooperative {NOMA}-Based {mmWave} Networks with Fluid Antenna System ({FAS}) Receivers},
  year={2023},
  doi={10.1109/BalkanCom58402.2023.10167904}
}

@ARTICLE{11098630,
  author={J. Huangfu and Z. Song and T. Hou and A. Li and Y. Liu and A. Nallanathan and K.-K. Wong},
  journal={IEEE Trans. Wireless Commun.},
  title={Performance Analysis of Fluid Antenna System Under Spatially-Correlated {Rician} Fading Channels},
  year={2026},
  volume={25},
  pages={1394--1407},
  doi={10.1109/TWC.2025.3590722}
}

@ARTICLE{11277276,
  author={P. Maurya and D. Dixit and A. K. Mishra and V. Bhatia and O. Krejcar},
  journal={IEEE Wireless Commun. Lett.},
  title={Performance Analysis of {RIS}-Assisted Fluid Antenna Systems Over {Rician} Fading},
  year={2026},
  volume={15},
  pages={910--914},
  doi={10.1109/LWC.2025.3639907}
}

@ARTICLE{10694739,
  author={F. R. Ghadi and K.-K. Wong and F. J. López-Martínez and W. K. New and H. Xu and C.-B. Chae},
  journal={IEEE Trans. Wireless Commun.},
  title={Physical Layer Security Over Fluid Antenna Systems: Secrecy Performance Analysis},
  year={2024},
  volume={23},
  number={12},
  pages={18201--18213},
  doi={10.1109/TWC.2024.3463488}
}

@ARTICLE{9715064,
  author={Z. Chai and K.-K. Wong and K.-F. Tong and Y. Chen and Y. Zhang},
  journal={IEEE Commun. Lett.},
  title={Port Selection for Fluid Antenna Systems},
  year={2022},
  volume={26},
  number={5},
  pages={1180--1184},
  doi={10.1109/LCOMM.2022.3152451}
}

@ARTICLE{10375698,
  author={X. Lai and T. Wu and J. Yao and C. Pan and M. Elkashlan and K.-K. Wong},
  journal={IEEE Commun. Lett.},
  title={On Performance of Fluid Antenna System Using Maximum Ratio Combining},
  year={2024},
  volume={28},
  number={2},
  pages={402--406},
  doi={10.1109/LCOMM.2023.3348028}
}

@article{Atz25,
	author = {I. {Atzeni} and others},
	title = {Sub-{THz} communications: {Perspective} and results from the {Hexa-X-II} project},
	journal = {IEEE Open J. Commun. Soc.},
	volume = {6},
	pages = {7495--7540},
	year = {2025}}
\end{document}